\pdfoutput=1
\documentclass[sigconf]{acmart}

\usepackage{booktabs} 
\usepackage{tikz}
\usepackage{amsmath}
\usepackage{amsthm,enumitem}
\usepackage{subfigure}
\usepackage{graphicx,color}
\usepackage{soul}
\usetikzlibrary{automata, positioning}

\setcopyright{rightsretained}
\newcommand{\pr}{\text{P}}
\newcommand{\E}{\mathrm{E}}
\newcommand{\deq}{\triangleq}

\begin{document}
\title{A Transient Queueing Analysis under Time-varying Arrival and Service Rates for Enabling Low-Latency Services}

\author{Wonjun Hwang}
\affiliation{%
  \institution{UNIST}
  \streetaddress{UNIST-gil 50}
  \city{Ulsan, South Korea}  
  \postcode{44919}
}
\email{gwaka@unist.ac.kr}

\author{Yoora Kim}
\affiliation{%
  \institution{University of Ulsan}
  \streetaddress{Daehak-ro 93}
  \city{Ulsan, South Korea}  
  \postcode{44610}
}
\email{yrkim@ulsan.ac.kr}

\author{Kyunghan Lee}
\affiliation{%
  \institution{UNIST}
  \streetaddress{UNIST-gil 50}
  \city{Ulsan, South Korea}  
  \postcode{44919}
}
\email{khlee@unist.ac.kr}


\begin{abstract}
Understanding the detailed queueing behavior of a networking session is critical in enabling low-latency services over the Internet. Especially when the packet arrival and service rates at the queue of a link vary over time and moreover when the session is short-lived, analyzing the corresponding queue behavior as a function of time, which involves a transient analysis, becomes extremely challenging. In this paper, we propose and develop a new analytical framework that anatomizes the transient queue behavior under time-varying arrival and service rates even under unstable conditions. Our framework is capable of answering key questions in designing low-latency services such as the time-dependent probability distribution of the queue length; the instantaneous or time-averaged violation probability that the queue length exceeds a certain threshold; and the fraction of time during an interval $[0, t]$ at which the queue length exceeds a certain threshold. We validate our framework by comparing its prediction results over time with the statistical simulation results and confirm that our analysis is accurate enough. Our extensive demonstrations on the efficacy of the analytical framework in designing low-latency services reveal that its prediction ability for the transient queue behavior in diverse time-varying packet arrival and service patterns can be of a high practical value.    
\end{abstract}

%
%




\maketitle

\pdfoutput=1
\section{Introduction}
\label{sec:intro}

Low-latency guarantee has been arguably the most challenging mission in the Internet during the last two decades.  There have been a number of proposals to control and reduce the Internet latency such as IntServ~\cite{intserv1994}, DiffServ~\cite{diffserv1998}, AQM (active queue management) techniques~\cite{codel2012, pie2013}, and low-latency TCPs~\cite{dctcp2010, drwa2016, bbr2016}. However even with such efforts, most real-time Internet applications of today including video conferencing and online gaming are still suffering from insufficient latency performance. In the near future, it seems like that a much wider range of applications will be subject to latency problems given that future Internet visions such as \textit{tactile Internet}~\cite{tactile2014} and 5G designs~\cite{5Gtactile2016, 5Gconcerns2014} forecast upcoming unprecedented demands for emerging real-time services such as AR (augmented reality), VR (virtual reality), and MR (mixed reality, a mixture of AR and VR~\cite{mr2014}). One of the last resort solutions we have, so called over-provisioning, may alleviate the pain, but it will be still largely insufficient to satisfy latency critical services like \textit{remote surgery} and \textit{telepresence}. 

We find that the major difficulty in controlling the latency in the Internet lies in its unpredictability, where the unpredictability mainly comes from the nature of packet multiplexing and queueing. Multiple observations made at large-scale data centers~\cite{incast2008, hull2012} show that comparing to other latency components such as propagation, transmission, and processing latencies, the queueing latency is fluctuating the most and hard to be tamed. More recent observations~\cite{bufferbloat2011, drwa2016} on the latency explosion problem, named \textit{bufferbloat}, confirm that when queueing becomes excessive, it can dramatically exacerbate the latency performance. To this end, we focus on understanding the queueing latency more deeply in this work. More specifically, we give our attention to developing an advanced analytical framework for queueing analysis. 


\begin{figure}[!t]
	\centering
	\includegraphics[width=0.85\columnwidth]{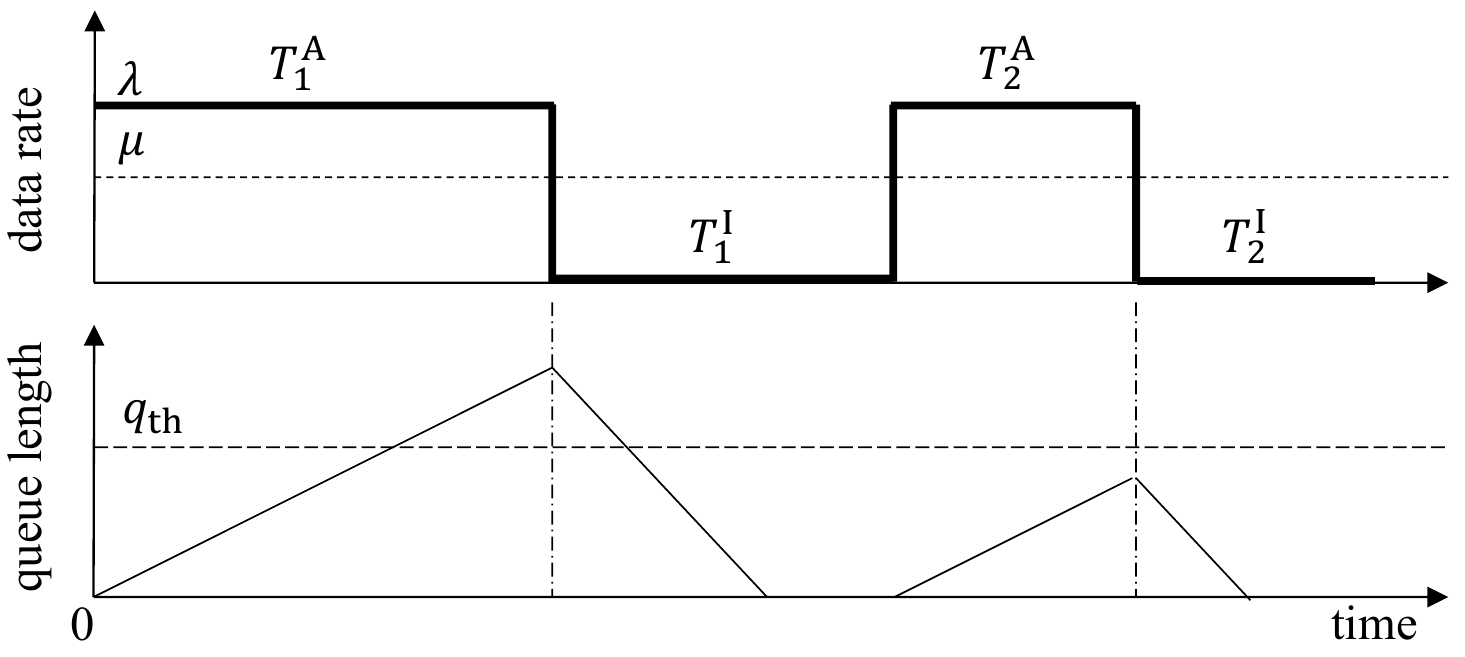}
    \caption{An example of a transient queueing system where the incoming data rate higher than the service rate ($\lambda > \mu$) persists during the limited time periods \textcolor{black}{$T_{1}^{\text{A}}$ and $T_{2}^{\text{A}}$}. It is possible for our analytical framework to answer the probability \textcolor{black}{that an incoming packet who arrives at time $t$} sees the queue length being larger than a threshold value $q_{\text{th}}$.}	
    \label{fig:problem}
\end{figure}

In order to widen the understanding, our analytical framework aims at giving detailed prediction of the queueing behavior as a function of time, i.e., transient queueing behavior, especially when the input traffic to a queueing system is non-stationary. 
The transiency in queueing of our interest further includes the situation where the rate of input traffic (i.e., arrival rate) intermittently exceeds the available service rate of the system.  We find that having detailed understanding of the queueing behavior for the transient queueing system is particularly important because most latency critical services may practically show such a traffic pattern and their resulting latency performance is currently under-explored. For instance, when a surgical robot needs to be controlled remotely in a mixed reality manner, the surgeon in operation may intermittently generate operational commands and receive feedback while the real-time multimedia traffic persists in the background, which makes the aggregated traffic for this session transient. A quick example of such a scenario is depicted in Figure~\ref{fig:problem}, where $\lambda$ and $\mu$ denote the arrival rate and the service rate of a system, respectively, and the transiency of $\lambda$ presents in the intermittent traffic arrival periods: \textcolor{black}{$T_{1}^{\text{A}}$ and $T_{2}^{\text{A}}$}. Even in this simplistic scenario, basic analytical questions such as \textit{How much the per-packet queueing latency will be?} and \textit{How much the per-packet queuing latency will violate a certain latency requirement?} are hard to be answered by most conventional queueing techniques. Given that the practical latency critical services in the near future may show more complicated mixture of diverse real-time traffic and the session duration of such services can also be various rather than being persistent, developing an analytical framework that takes the transiency into account is essential. 

The analytical method in our framework is different from conventional queueing techniques such as MMPP (Markov modulated Poisson process), EMC (Embedded Markov chain), or Phase-type distributions. For the detailed analysis of the transient queueing system to which any steady-state analysis is not applicable, we adopt a transient queue analysis technique introduced in~\cite{transient1993}, which first gave expressions for the probabilistic queue length behavior in transient queueing systems. We find that our framework performing a judicious numerical analysis for the mathematical expressions is capable of demystifying a wide range of challenging queueing latency problems. 
\pdfoutput=1
\section{Related Work}
\label{sec:related}

We here introduce several recent studies in two separate research directions: queueing latency reduction algorithms and queueing latency analysis techniques. Note that there are many other studies aiming at reducing latency of the Internet including new architecture and protocol designs, but in order to keep our problem focused, we intentionally narrow down the scope of related work. 

\subsection{Queueing latency reduction algorithms}
The first systematic observation of bufferbloat phenomenon made at~\cite{bufferbloat2011} revealed that all the TCPs (transmission control protocols) using packet loss as a congestion indicator are subject to excessive queueing latency, where the excessiveness is determined by the buffer size of the queue in the bottleneck link. For a loss-based TCP to experience packet losses in a drop-tail queue, such a TCP is designed to always fill up the queue. Recalling the basics of queueing theory, this operation makes a queue being operated at the work-conserving regime (i.e., with non-zero packets queued all the time), thus achieving the maximum throughput, but at the same time substantially increasing the queueing latency. TCP algorithms for receivers, DRWA~\cite{drwa2016} and senders, BBR~\cite{bbr2016} tackle this latency problem by limiting the amount of queuing in the bottleneck link using the concept of BDP (bandwidth delay product) where the BDP is interpreted in concept as the minimum amount of per-RTT input traffic that achieves both the minimum latency as well as the maximum throughput when the input and output of a queue is nearly deterministic. In a realistic situation where randomness exists, two performance metrics, latency and throughput, are partially in a trade-off relation, and thus achieving the best for latency and throughput together becomes much more complicated. For this, DRWA, BBR, and other recent congestion control algorithms such as VERUS~\cite{verus2015} and TCP Ex-Machina~\cite{exmachina2013} design their own congestion window (CWND) control schemes that adaptively suppress the queueing amount for lowering queueing latency while sustaining the throughput. However, because these algorithms commonly use empirical adjustment strategy which repeats a two-step procedure {\em Adjust and Observe}, they can improve the latency performance but cannot guarantee it. In order for a transport layer protocol to guarantee the latency performance, a prediction method is required before taking an adjustment action, which can be realized by a latency modelling and analysis. 

\subsection{Queueing latency analysis techniques}
Analyzing latencies of a network such as per-hop latency and end-to-end latency has been tried in many ways. While the most of literature focused on analyzing the expected latency, there existed other approaches. Bisti et al.~\cite{e2etandem2012} provided a numerical analysis of the worst-case end-to-end delay bound for a tandem of queues using a network calculus technique. Baik and Nadakuditi~\cite{tandemQ2014} analyzed the sojourn time distributions of the last packet in a packet batch in each queue of a tandem of queues. Rather than studying several characteristics of the end-to-end latency, there were studies anatomizing a much more detailed latency behavior of a packet in a queue from which people started to devise a new technique called \textit{transient analysis}. Abate and Whitt~\cite{transient1987} provided the first transient analysis for the M/M/1 queue by which the latency distribution of each individual packet is understood. More systematic transient analyses for the M/M/1 queue and for the M/G/1 queue are made later by Leguesdron et al.~\cite{transient1993} and Wang et al.~\cite{transient2008}, respectively. Although the transient analysis can be extremely useful in precisely predicting the queueing delay of a network, this technique has been underexploited in the networking community. Also, the adaptation of the transient analysis to a more practical scenario involving a queueing system where the packet arrival rates and service rates are time-varying has been under-explored.


\pdfoutput=1
\section{System Description}
\label{sec:system}

In this section, we describe the settings of the queueing system that we analyze throughout this paper. Then, we interpret the system in the perspective of low-latency Internet applications. Lastly, we clarify the problem to solve in this paper.

\subsection{System Model}
\label{sec:system:model}
We target to analyze a generalized single-server queueing system in which its packet arrival rate and service rate can vary over time. The main assumptions regarding the queue modelling are summarized as follows: 
\begin{itemize}[leftmargin=*]
\item (A1) Packets arrive at the queue according to a non-homogeneous Poisson process with time-varying rate $\lambda(t)$ (packets/sec).
\item (A2) The size of a packet is exponentially distributed with mean $L$ (bits). Without loss of generality, we set $L=1$. 
\item (A3) The server uses a first-in-first-out policy to serve each packet and supports service rate $\mu(t)$ (bits/sec).
\item (A4) The queue has infinite capacity. 
\end{itemize}

\begin{figure}[!t]
	\centering
	\includegraphics[width=0.85\columnwidth]{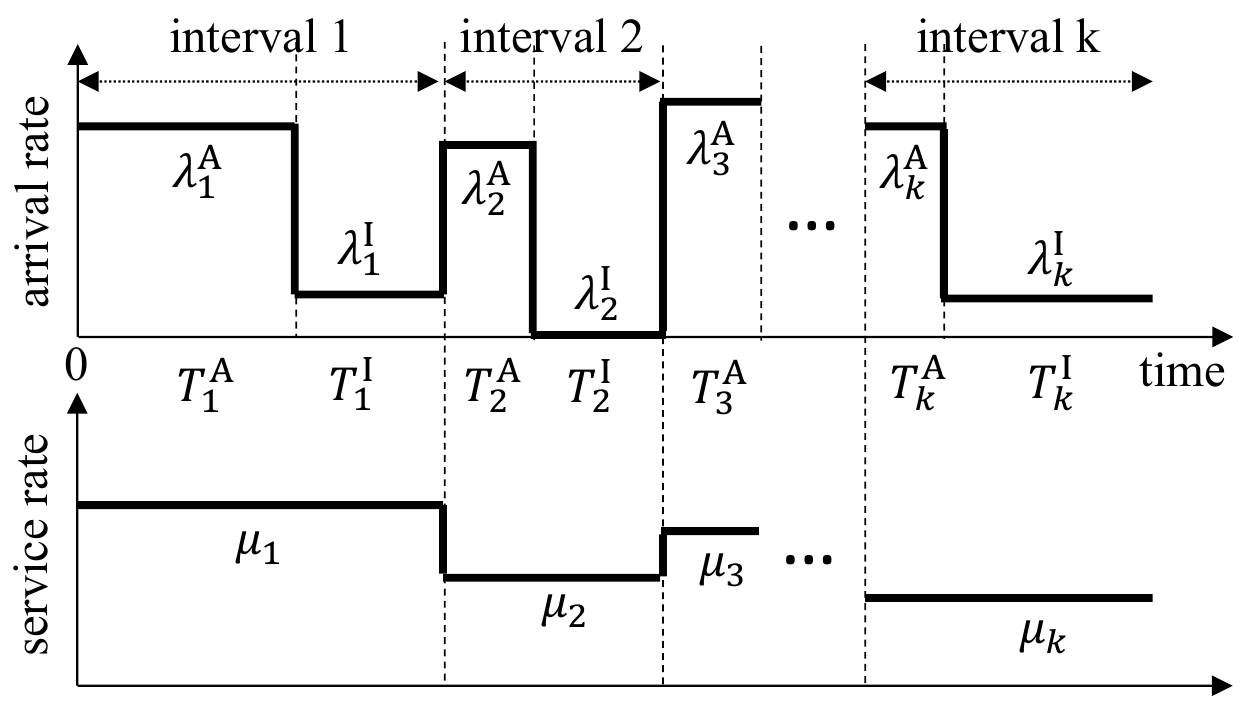}
	\caption{The arrival and service rate behaviors of a generalized single-server queueing system of our interest. The arrival rate changes at every active or inactive duration of each interval while the service rate changes at every interval.}	
    \label{fig:system}
\end{figure}

In the following, we detail how the rate parameters $\lambda(t)$ and $\mu(t)$ vary over time. The time axis is divided into \emph{intervals}, and each interval consists of an \emph{active} subinterval followed by an \emph{inactive} subinterval (see Figure~\ref{fig:system}). The lengths of the active and inactive subintervals of the $k$th interval ($k=1,2,3,\ldots$) are denoted by $T_k^{\text{A}}$ (sec) and $T_k^{\text{I}}$ (sec), respectively, and they can take different values across~$k$. Hence, the $k$th interval corresponds to the time interval $[S_{k}, S_{k+1})$ where $S_1 = 0$, 
\begin{align*}
    S_k &= \sum_{n=1}^{k-1}(T_n^{\text{A}}+T_n^{\text{I}}), \qquad k = 2, 3, 4, \ldots,
\end{align*}
and the interval $[S_{k}, S_{k+1})$ is divided into two subintervals: active subinterval $[S_{k}, S_{k}+T_k^{\text{A}})$ and inactive subinterval $[S_{k}+T_k^{\text{A}}, S_{k+1})$.

We assume that the arrival rate $\lambda(t)$ and the service rate $\mu(t)$ follow step functions given by
\begin{align*}
\begin{split}
\lambda(t) &= \begin{cases}
    \lambda_k^{\text{A}} & t \in [S_{k}, S_{k}+T_k^{\text{A}}),\\
    \lambda_k^{\text{I}} (\ll \lambda_k^{\text{A}}) & t\in [S_{k}+T_k^{\text{A}}, S_{k+1}),
\end{cases} \\
\mu(t) &= \mu_k \hspace{5.5mm} \quad \quad \quad   t \in [S_{k}, S_{k+1}).
\end{split}
\end{align*}
That is, the arrival rate $\lambda(t)$ changes from subinterval to subinterval, while the service rate $\mu(t)$ changes from interval to interval. \textcolor{black}{Accordingly, the \emph{switching point} (i.e., the instant at the beginning of each active or inactive subinterval) is given in sequence as 
$$S_1 \leq  S_1 + T_1^{\text{A}} \leq S_2 \leq S_2 + T_2^{\text{A}} \leq \cdots.$$}
We clarify here that the rate of input traffic $\lambda(t)$ is allowed to exceed the available service rate $\mu(t)$ during some subintervals so that the queueing system is possibly \emph{unstable intermittently}. We aim at developing an analytical framework that can handle even such a queue bubbling case, 
which is known intractable by any steady-state analysis. 

\begin{figure}[!t]
	\centering
	\includegraphics[width=0.85\columnwidth]{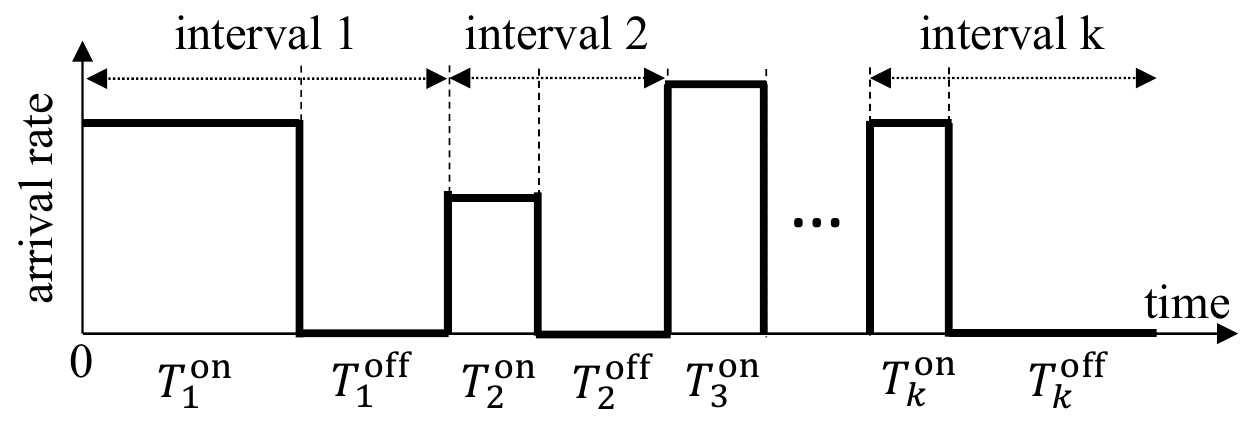}
    \caption{A depiction of the packet arrival behavior over time in the on and off traffic generation model.}	
    \label{fig:caseonoff}
\end{figure}

\begin{figure}[!t]
	\centering
	\includegraphics[width=0.85\columnwidth]{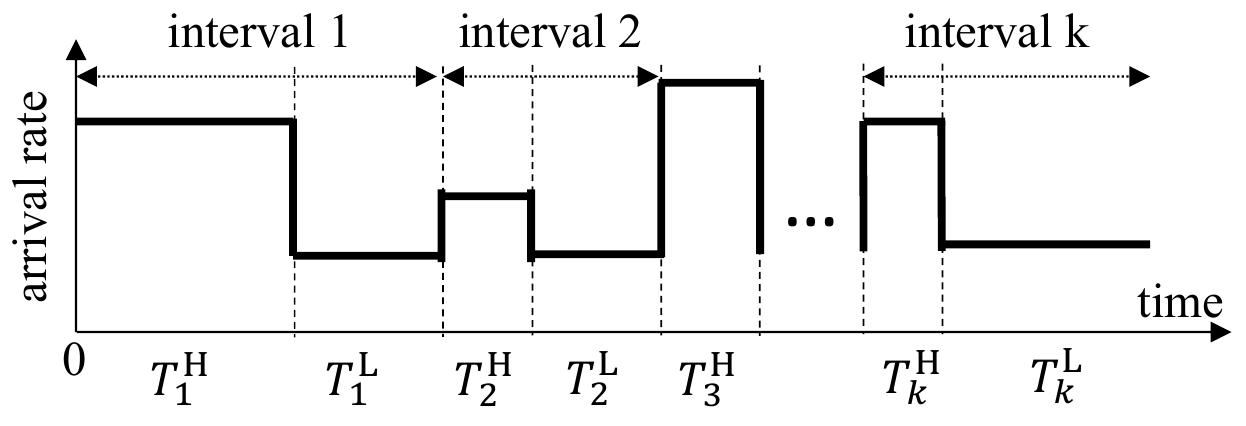}
    \caption{A depiction of the packet arrival behavior over time in the high and low traffic generation model.}	
    \label{fig:casehighlow}
\end{figure}

\begin{figure}[!t]
	\centering
	\includegraphics[width=0.85\columnwidth]{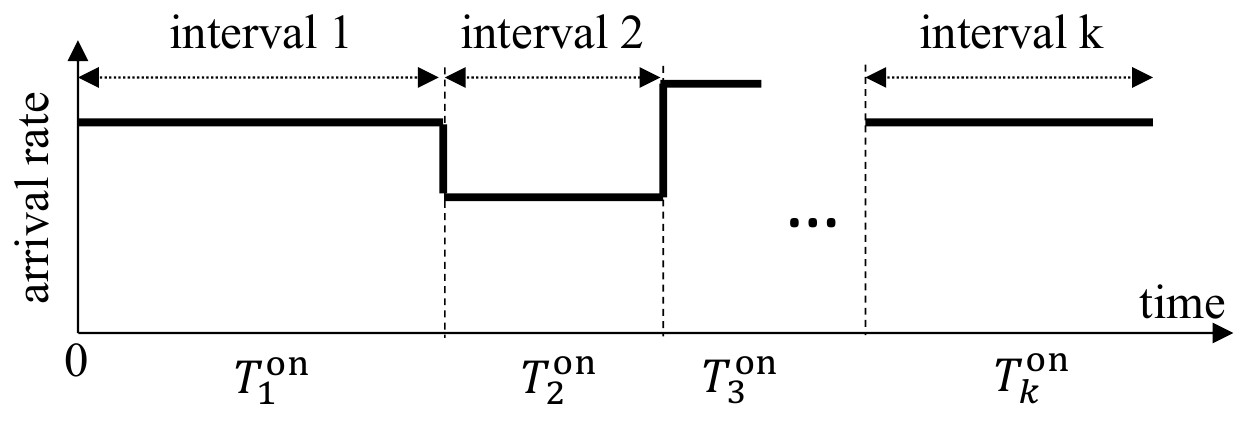}
    \caption{A depiction of the packet arrival behavior over time in the restless traffic generation model.}	
    \label{fig:caserestless}
\end{figure}

\subsection{Applications}
Our analytical framework can be useful in many areas, but in this work we give our special attention to low-latency Internet applications, where the question, \textit{how much portion of the packets of a session can arrive within a latency bound?}, is most critical. In order to emulate general traffic generation behaviors of low-latency applications, we propose three simplified \textcolor{black}{$\lambda(t)$} and $\mu(t)$ evolution models and perform in-depth studies for such models. 

\smallskip
\noindent{\bf{On-Off Traffic Model:}}
With this model, we emulate the case where a low-latency service user gives timely control toward a remote machine such as a Drone or a self-driving vehicle. Because the control packets are generated intermittently and there exist alternating periods of no control, the packet arrival toward the queueing system from the controlling user resembles alternating on and off periods as shown in Figure~\ref{fig:caseonoff}. We here note that even when there is no other competing flows in the queueing system, this intermittent packet burst itself builds up the queue and makes a portion of succeeding packets to violate the latency requirement. This is so called, \textit{self-induced queueing}, which is well analyzed by our framework. We also note that the situation where there is no competing flow in a queueing system frequently happens in the cellular base station because most modern base stations create a virtual individual queue, called \textit{bearer}, upon reception of a flow toward a cellular device (i.e., a remote machine in this scenario)~\cite{drwa2016}. The virtual queues in a base station do not interfere each other, so the queueing in the virtual queue is mostly self-induced. 

\smallskip
\noindent{\bf{High-Low Traffic Model:}}
With this model, we emulate the case where a low-latency service user intermittently creates control toward a remote machine while a background media stream such as video or audio session persists over time. We expect that this pattern of traffic generation will frequently occur in the AR or MR type of services, in which the users need to transmit their multimedia information along with their control information with latency guarantee so that the server system can manipulate or augment the multimedia data and return the data to the user with nearly unnoticeable delay. We capture such a persistent background stream with the rate of $\lambda^{L}_k$ and its merged stream with intermittent control packets as $\lambda^{H}_k$, where $H$ and $L$ stand for high and low. This model is depicted in Figure~\ref{fig:casehighlow}. 

\smallskip
\noindent{\bf{Restless Traffic Model:}}
With this traffic model, more generalized low-latency services are emulated. By letting all inactive subintervals be zero (i.e., active subinterval occupies each interval), the arrival rate is modeled to be freely changing at every interval \textcolor{black}{as shown in Figure~\ref{fig:caserestless}}. Because now the service rate change is synchronized with the arrival rate, it becomes simpler to capture an arbitrary available bandwidth fluctuation by regarding each interval in the time scale of environmental change. Low-latency services between machine to machine where the input traffic intermittency may not exist can be of representative examples of this model. Moreover, general low-latency services that do not fall in the categories of on-off or high-low traffic models can be emulated with this model.

\subsection{Problem Statement}
\label{sec:system:problem}
Let $Q(t)$ be a random variable denoting the number of packets in the system at time~$t$.\footnote{$Q(t)$ is the sum of the number of packets in the queue and that in the server. The number of packets in the server is up to 1. Hence, for simplicity, we shall call $Q(t)$ the queue length at time $t$ unless confusion arises. } For the queueing system described in Section~\ref{sec:system:model}, we aim to find the probability distribution of the queue length $Q(t)$ as a function of time $t$. That is, we solve for $p_n(t)$, called the \emph{transient solution}, where
\begin{align}\label{eqn:transient_sol}
p_n(t) \deq \pr\{Q(t) = n\}, \qquad n = 0, 1,2,\ldots.    
\end{align}
Knowing the transient solution $[p_n(t), n \geq 0]$ enables us to fully characterize the statistical properties of the queueing system including the average queue length, the queueing delay, and the buffer overflow probability as a function of time.

Above all, in the perspective of low-latency networking, we give our special attention to the probability that the queue length at time $t$ is larger than a given threshold $q_{\text{th}}$. To formalize, we define 
\begin{align}\label{eqn:metric:1}
    V(t) &\deq \pr\{Q(t) > q_{\text{th}}\},
\end{align}
and call $V(t)$ the \emph{violation probability} throughout the paper. The metric $V(t)$ is interpreted as the probability for an incoming packet who arrives at time~$t$ to see the queue length being larger than $q_{\text{th}}$. 

Another important performance metric is the average value of $V(\cdot)$ on the interval $[0, t]$, which we define as 
\begin{align}\label{eqn:metric:2}
    \bar{V}(t) &\deq  \frac{1}{t}\int_{0}^{t} V(s) \, \mathrm{d}s.
\end{align}
We call $\bar{V}(t)$ the \emph{time-averaged violation probability}. 
Note that $V(s) = \E[ 1_{\{Q(s) > q_{\text{th}}\}}]$, where $1_{\{\cdot\}}$ is the indicator function. Accordingly, the integral in (\ref{eqn:metric:2}) is equal to 
\begin{align*}
\int_{0}^{t} V(s) \, \mathrm{d}s =\int_{0}^{t} \E[ 1_{\{Q(s) > q_{\text{th}}\}}] \, \mathrm{d}s = \E[\int_{0}^{t} 1_{\{Q(s) > q_{\text{th}}\}} \, \mathrm{d}s],
\end{align*}
and thus we can express $\bar{V}(t)$ in fractional form as 
\begin{align*}
   \bar{V}(t) = \frac{\E[\int_{0}^{t} {1}_{\{Q(s) > q_{\text{th}}\}} \, \mathrm{d}s ]}{t}.
\end{align*}
In this fraction, the numerator indicates the average total duration of time within the interval $[0, t]$ at which the queue length exceeds the threshold $q_{\text{th}}$, whereas the denominator indicates the duration of $[0, t]$. \textcolor{black}{Therefore, we can interpret $\bar{V}(t)$ as the fraction of time being overflowed during the interval $[0,t]$.} Such metric is especially useful in estimating the number of packets delivered within a given latency bound among the packets transmitted during $[0,t]$.

We note that our key performance metrics defined in (\ref{eqn:metric:1}) and (\ref{eqn:metric:2}) can be obtained from the transient solution $p_n(t)$ in~(\ref{eqn:transient_sol}) as follows:
\begin{align*}
    V(t) &= 1 - \sum_{n = 0}^{q_{\text{th}}} p_n(t), \\
    \bar{V}(t) &= 1 -\frac{1}{t} \sum_{n = 0}^{q_{\text{th}}} \int_{0}^{t}p_n(s)\, \mathrm{d}s.
\end{align*}
\textcolor{black}{For simplicity, we use $\bar{V}$ and $\bar{V}(t)$ interchangeably if~$t$ refers to the end point of the group of intervals that is under investigation.} 

\pdfoutput=1
\section{Analytical Framework}
\label{sec:analysis}
In this section, we develop an analytical framework for finding the transient solution $p_n(t)$ of the problem stated in Section~\ref{sec:system:problem}. We first explain the technical approach residing in our framework. 

\subsection{Technical Approach}
\begin{figure}[!t]
\begin{center}
\begin{tikzpicture}
\node[state]             (s0) {0};
\node[state, right=of s0] (s1) {1};
\node[state, right=of s1] (s2) {2};
\node[draw=none, right=of s2] (s3)  {$\cdots$};

\draw[every loop]
    (s0) edge[bend left, auto=left] node {\(\lambda(t)\)} (s1)
    (s1) edge[bend left, auto=left] node {\(\mu(t)\)} (s0)
    (s1) edge[bend left, auto=left] node {\(\lambda(t)\)} (s2)
    (s2) edge[bend left, auto=left] node {\(\mu(t)\)} (s1)
    (s2) edge[bend left, auto=left] node {\(\lambda(t)\)} (s3)
    (s3) edge[bend left, auto=left] node {\(\mu(t)\)} (s2);
\end{tikzpicture}
\end{center}
\caption{Transition diagram of the process $\{Q(t); t\geq 0\}$}
\label{Fig:TD}
\end{figure}
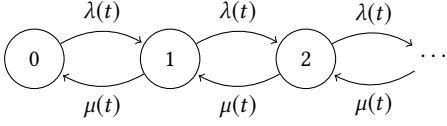

Our analysis is built upon the assumptions presented in Section~\ref{sec:system:model}. The assumption (A1) results in 
\begin{align*}
\pr\{\text{0 arrival in } (t, t+\Delta t]\} &= 1-\lambda(t) \Delta t + o(\Delta t),\\
\pr\{\text{1 arrival in } (t, t+\Delta t]\} &= \lambda(t) \Delta t + o(\Delta t), 
\end{align*}
where $\Delta t$ is a sufficiently small value, and a function $f(\cdot)$ is $o(h)$ if $\lim_{h\to 0}\frac{f(h)}{h}=0$. The assumptions (A2) and (A3) yield
\begin{align*}
\pr\{\text{0 departure in } (t, t+\Delta t]\} &= 1-\mu(t) \Delta t + o(\Delta t), \\
\pr\{\text{1 departure in } (t, t+\Delta t]\} &= \mu(t) \Delta t + o(\Delta t). 
\end{align*}
It then follows that the number of packets in the system, $Q(t)$, evolves according to a birth-death type of a continuous-time Markov process whose transition diagram is depicted in Figure~\ref{Fig:TD}. From the Chapman-Kolmogorov forward differential-difference equations, we obtain
\begin{align}\label{eqn:governing}
\begin{split}
\dfrac{\mathrm{d} p_n(t)} {\mathrm{d} t} & = -(\lambda(t) \!+\! \mu(t))p_n(t) \!+\! \lambda(t) p_{n\!-\!1}(t) \!+\! \mu(t) p_{n
\!+\!1}(t),~~n \!\geq \!1, \\
\dfrac{\mathrm{d}  p_0(t)} {\mathrm{d} t}  & = -\lambda(t) p_0(t) + \mu(t) p_1(t).
\end{split}
\end{align}
Hence, the transient solution $p_n(t)$ is obtained by solving the set of equations (\ref{eqn:governing}). To this end, we first consider a preliminary case where
$\lambda(t) = \lambda$ and $\mu(t) = \mu$ for all $t\geq 0$. In this case, our problem reduces to finding the transient solution of an M/M/1 queue. Based on this preliminary analysis, we next derive a recursive formula for the queue length distribution at every switching point. Finally, by invoking the property that the rate parameters $\lambda(t)$ and $\mu(t)$ remain constant between two adjacent switching points, we find an expression for $p_n(t)$ for an arbitrary time $t$ (see Theorem~1). 

\subsection{Preliminary}
\label{sec:analysis:pre}
In this section, we find the solution of the equation~(\ref{eqn:governing}) when $\lambda(t) = \lambda \geq 0$ and $\mu(t) = \mu>0$ for all $t\geq 0$. 
We provide separate analyses for two branches where $\lambda > 0$ or $\lambda = 0$. In both branches, we suppose $Q(0) = j$, i.e., the initial condition of the equation (\ref{eqn:governing}) is imposed as 
\begin{align}\label{eqn:governing:initial}
    p_n(0) = \begin{cases}1 & n = j, \\
    0 & n \neq j.
    \end{cases}
\end{align}
First, we consider the case $\lambda >0$. Then, the solution of the equation~(\ref{eqn:governing}) subject to the condition (\ref{eqn:governing:initial}) is known \textcolor{black}{in~\cite{gross1998fundamentals} as:}\footnote{The main idea is to take double transforms for the set of equations (\ref{eqn:governing}) both on the discrete variable $n$ and on the continuous variable $t$ to obtain a single equation, and then take double inverse transforms for the solution. For details, refer to~\cite{gross1998fundamentals}.}
\begin{align}\label{eqn:mm1:1}
\begin{split}
    p_n(t) &= e^{-(\lambda + \mu)t }\rho^{ \frac{n-j}{2} } I_{n-j}(2\mu\sqrt{\rho} t) + \rho^{\frac{n-j-1}{2}}I_{n+j+1}(2\mu\sqrt{\rho}t)\\ & \qquad +
(1 -\rho)\rho^n \sum_{s=n+j+2}^{\infty} { \rho^{-\frac{s}{2}}I_s(2\mu\sqrt{\rho}t) },
\end{split}
\end{align}
where \(\rho = \frac{\lambda}{\mu}\,(>0)\), and $I_s(x)$ is the modified Bessel function of the first kind and is of order $s$ defined by
\begin{align*}
I_s(x) \deq \sum_{m=0}^{\infty}{\dfrac{ 1} { (s+m)!m! } \left(\frac{x}{2}\right)^{s+2m}}.
\end{align*}
Next, we consider the case $\lambda = 0$. Then, the equation (\ref{eqn:governing}) describes a pure death process in which $Q(t)$ decreases to $Q(t)-1$ with rate $\mu$ when $Q(t) \geq 1$. Hence, we obtain 
\begin{align}\label{eqn:mm1:2}
\begin{split}
    p_n(t) = \begin{cases}
        \displaystyle \sum_{s=j}^{\infty}e^{-\mu t}\frac{(\mu t)^{s}}{s!} &  n = 0, \\
    \displaystyle  e^{-\mu t}\frac{(\mu t)^{j-n}}{(j-n)!} & 0 < n \leq j, \\
    0 & n >j.
    \end{cases}
    \end{split}
\end{align}
Combining the results in (\ref{eqn:mm1:1}) and (\ref{eqn:mm1:2}), we have Lemma~1.
\smallskip

\noindent \textbf{Lemma~1 (Transient solution when $\lambda(t)= \lambda$ and $\mu(t)=\mu$).} \\
The solution of the following initial value problem 
\begin{align*}
\dfrac{\mathrm{d} p_n(t)} {\mathrm{d}t} & = -(\lambda + \mu)p_n(t) + \lambda p_{n-1}(t) + \mu p_{n
+1}(t),\quad  n\geq 1,\\
\dfrac{\mathrm{d} p_0(t)} {\mathrm{d}t}  & = -\lambda p_0(t) + \mu p_1(t),\\
    p_n(0) &= \begin{cases}1 & n = j, \\
    0 & n \neq j,
    \end{cases}
\end{align*}
is given by 
$$p_n(t) = G_{j,n}(t;\lambda,\mu),$$ where
$G_{j,n}(t;\lambda,\mu)$ is given at the top of the next page. 

\noindent \textbf{Proof.} See the derivation from  (\ref{eqn:governing:initial}) to (\ref{eqn:mm1:2}).  \hfill $\blacksquare$
\smallskip

\begin{figure*}[!t]
\normalsize
\begin{align*}
\begin{split}
    G_{j,n}(t;\lambda,\mu) \!\deq\!
    \begin{cases}
     e^{-(\lambda + \mu)t }  \!  \rho^{ \frac{n-j}{2} } I_{n-j}(2\mu \! \sqrt{\rho} t)\! + \! \rho^{\frac{n-j-1}{2}}I_{n+j+1}(2\mu\sqrt{\rho}t) \!+\!
(1 \!-\!\rho)\rho^n \displaystyle \sum_{s=n+j+2}^{\infty} { \rho^{-\frac{s}{2}}I_s(2\mu\sqrt{\rho}t) }
    & \lambda>0, \\
    \displaystyle \sum_{s=j}^{\infty}e^{-\mu t}\frac{(\mu t)^{s}}{s!} &    \lambda=0, \,  n = 0, \\
   \displaystyle e^{-\mu t}\frac{(\mu t)^{j-n}}{(j-n)!} & \lambda=0, \,0 < n \leq j, \\
    0 &  \lambda=0, \,n >j.
\end{cases}
\end{split}
\end{align*}
\hrulefill
\end{figure*}

\noindent \textbf{Remark~1.} $G_{j,n}(t;\lambda,\mu)$ is a key function in constructing our analytical framework. We remark that in $G_{j,n}(t;\lambda,\mu)$ the argument $t$ indicates the elapsed time from the initial time point; the subscripts $j,n$ imply state transition from $j$ to $n$ after the amount of time $t$ is elapsed; and the parameters $\lambda, \mu$ are the arrival rate and the service rate, respectively.

\subsection{Analytical Framework}
In this section, we present our analytical framework. We begin with Lemma~2 which characterizes the transition probability of the queue length during the $k$th interval ($k=1,2, 3, \ldots$).

\smallskip
\noindent \textbf{Lemma~2 (Transition probability during the $k$th interval).} \\
(i) For any $t_1$ and $t_2$ such that $S_k \leq t_1 \leq t_2 \leq S_k + T_k^{\text{A}}$, the queue-length transition from $t_1$ to $t_2$ is governed by
$$\pr\{Q(t_2) = n\,|\, Q(t_1)=j\} = G_{j,n}(t_2-t_1; \lambda_k^{\text{A}}, \mu_k).$$
(ii) For any $t_1$ and $t_2$ such that $S_k + T_k^{\text{A}} \leq t_1 \leq t_2 \leq S_{k+1}$, the queue-length transition from $t_1$ to $t_2$ is governed by
$$\pr\{Q(t_2) = n\,|\, Q(t_1)=j\} = G_{j,n}(t_2-t_1; \lambda_k^{\text{I}}, \mu_k).$$

\noindent \textbf{Proof.} If $S_k \leq t_1 \leq t_2 \leq S_k + T_k^{\text{A}}$, then $[t_1, t_2]$ is a subset of the $k$th active subinterval. Hence, for any $t\in [t_1, t_2]$, the equation~(\ref{eqn:governing}) reduces to 
\begin{align*}
\dfrac{\mathrm{d} p_n(t)} {\mathrm{d}t} & = -(\lambda_k^{\text{A}} + \mu_k)p_n(t) + \lambda_k^{\text{A}} p_{n-1}(t) + \mu_k p_{n
+1}(t), \quad n\geq 1, \\
\dfrac{\mathrm{d} p_0(t)} {\mathrm{d}t}  & = -\lambda_k^{\text{A}} p_0(t) + \mu_k p_1(t).
\end{align*}
The condition $Q(t_1) = j$ translates into 
\begin{align*}
p_n(t_1) &= \begin{cases}1 & n = j, \\
    0 & n \neq j.
    \end{cases}
\end{align*}
Therefore, using the change of variables $\tilde{t} = t-t_1$ and then applying Lemma~1 with $\lambda = \lambda_k^{\text{A}}$ and $\mu = \mu_k$, we obtain
$$\pr\{Q(t_2) = n\,|\, Q(t_1)=j\} = G_{j,n}(t_2-t_1; \lambda_k^{\text{A}}, \mu_k),$$
which proves Lemma~2~(i).

If $S_k + T_k^{\text{A}} \leq t_1 \leq t_2 \leq S_{k+1}$, then $[t_1, t_2]$ is a subset of the $k$th inactive subinterval. Following the approach used in the proof of Lemma~2~(i), we can prove Lemma~2 (ii). Due to similarity, we omit the details. \hfill $\blacksquare$
\smallskip

Based on Lemma~1, we next find the transient solution $p_n(t)$ at every switching point $S_1 \leq  S_1 + T_1^{\text{A}} \leq  S_2 \leq  S_2+T_2^{\text{A}} \leq S_3\leq  \cdots$, which we denote by
\begin{align*}
A_{k,n} &\deq p_n(S_k), \\
I_{k,n} &\deq  p_n(S_k+T_k^{\text{A}}).
\end{align*}
By the law of total probability, we have 
\begin{align}\label{eqn:initial-1}
I_{k,n} 
&= \sum_{j=0}^{\infty}  \pr\{Q(S_k)=j, Q(S_k+T_k^{\text{A}}) = n\} \nonumber \\
&= \sum_{j=0}^{\infty}  \pr\{Q(S_k)=j\} \cdot \pr\{Q(S_k+T_k^{\text{A}}) = n\,|\, Q(S_k) = j\}.
\end{align}
From Lemma~2~(i), the transition probability in the right-hand side of (\ref{eqn:initial-1}) is obtained by 
\begin{align}\label{eqn:initial-2}
\pr\{Q(S_k+T_k^{\text{A}}) = n\,|\, Q(S_k) = j\} &=  G_{j,n}(T_k^{\text{A}}; \lambda_k^{\text{A}}, \mu_k).
\end{align}
Substituting (\ref{eqn:initial-2}) into (\ref{eqn:initial-1}) gives
\begin{align}\label{eqn:initial-3}
I_{k,n} &=  \sum_{j=0}^{\infty}A_{k,j} \cdot G_{j,n}(T_k^{\text{A}}; \lambda_k^{\text{A}}, \mu_k).
\end{align}
Hence, if we are given the distribution $[A_{k,n}, n \geq 0 ]$, we can compute the distribution $[I_{k,n}, n \geq 0 ]$ using the formula~(\ref{eqn:initial-3}). Applying a similar argument, we obtain 
\begin{align}\label{eqn:initial-4}
A_{k+1,n} &= \sum_{j=0}^{\infty}  \pr\{Q(S_{k}+T_{k}^{\text{A}})=j, Q(S_{k+1}) = n\} \nonumber \\
&= \sum_{j=0}^{\infty}  \pr\{Q(S_{k}+T_{k}^{\text{A}})=j\} \cdot \pr\{Q(S_{k+1}) = n\,|\, Q(S_{k}+T_{k}^{\text{A}}) = j\}\nonumber \\
&= \sum_{j=0}^{\infty} I_{k,j}\cdot G_{j,n}(T_{k}^{\text{I}}; \lambda_{k}^{\text{I}}, \mu_{k}),
\end{align}
which implies that, if we are given the distribution  $[I_{k,n}, n \geq 0 ]$, we can compute the distribution $[A_{k+1,n}, n \geq 0 ]$. Therefore, by mathematical induction, we can find $[p_n(t), n \geq 0]$ for any switching point $t$, provided that the initial distribution $[p_n(0), n \geq 0 ]$ is given. The result is summarized in Lemma~3. 

\smallskip
\noindent \textbf{Lemma~3 (Transient solution at switching point).} \\ 
Let $p(t) = [p_n(t), n\geq 0]$. Then, for $k=1,2,3,\ldots$, we have
\begin{align*}
p(S_k) &= p(0)  \prod_{s=1}^{k-1}  \left( G_s^{\text{A}} \cdot  G_s^{\text{I}} \right), \\
p(S_k+T_k^{\text{A}}) &= p(0)  \prod_{s=1}^{k-1}  \left( G_s^{\text{A}} \cdot  G_s^{\text{I}} \right) G_k^{\text{A}},
\end{align*}
where $G_s^{\text{A}}$ and $G_s^{\text{I}}$ are matrices defined by 
\begin{align*}
    G_s^{\text{A}} & \deq \big[G_{i,j}(T_s^{\text{A}}; \lambda_s^{\text{A}}, \mu_s)\big]_{i,j\geq 0},\\
    G_s^{\text{I}} & \deq \big[G_{i,j}(T_s^{\text{I}}; \lambda_s^{\text{I}}, \mu_s)\big]_{i,j\geq 0}. 
\end{align*}
When $k=1$, we adopt the convention $\prod_{s=1}^{0}  \left( G_s^{\text{A}} \cdot  G_s^{\text{I}} \right)  = E$, where $E$ is an identity matrix.

\noindent \textbf{Proof.} The formulas (\ref{eqn:initial-3}) and (\ref{eqn:initial-4}) hold for all $n=0,1,2,\ldots$. Hence, we can represent them in matrix form as follows:
\begin{align*}
    p(S_k + T_k^{\text{A}}) &= p(S_k)\cdot G_k^{\text{A}}, \qquad k = 1, 2, 3,\ldots,\\
    p(S_{k+1}) &= p(S_k+T_k^{\text{A}})\cdot G_k^{\text{I}}, \qquad k = 1, 2, 3,\ldots.
\end{align*}
Since $p(S_1) = p(0)$, we have Lemma~3. \hfill $\blacksquare$
\smallskip

The distributions $p(S_k)$ and $p(S_k+T_k^{\text{A}})$ in 
Lemma~3 can be interpreted as the initial distributions of the queue length for the $k$th active and inactive subintervals, respectively, as the switching points are located at the beginning of each subinterval. Hence, using Lemma~2 on the transition probability and Lemma~3 on the initial distribution, we can now derive a formula for $p_n(t)$ for an arbitrary time $t\geq 0$. 

We first consider the case when $t$ is in the active duration of the $k$th interval, i.e., $t \in [S_k,  S_k + T_k^{\text{A}})$. By conditioning on the queue length at the beginning of the $k$th active subinterval, we have
\begin{align}\label{eqn:final-1}
p_n(t) &= \sum_{j=0}^{\infty} \pr\{Q(S_k)=j\} \cdot \pr\{Q(t) = n\,|\, Q(S_k) = j\}. 
\end{align}
Since $t \in [S_k,  S_k + T_k^{\text{A}})$, we can apply  Lemma~2~(i) with $t_1 = S_k$ and $t_2 = t$ to have 
\begin{align}\label{eqn:final-2}
\pr\{Q(t) = n\,|\, Q(S_k) = j\} =  G_{j,n}(t - S_k; \lambda_k^{\text{A}}, \mu_k).
\end{align}
Therefore, substituting (\ref{eqn:final-2}) into (\ref{eqn:final-1}) yields 
\begin{align}\label{eqn:final-3}
p_n(t) &=  \sum_{j=0}^{\infty } p_j(S_k) \cdot G_{j,n}(\Delta t; \lambda_k^{\text{A}}, \mu_k),
\end{align}
where $\Delta t = t - S_k$ is the elapsed time from the beginning of the $k$th active subinterval.

Next we consider the case when $t$ is in the inactive duration of the $k$th interval, i.e., $t \in [S_k + T_k^{\text{A}}, S_{k+1})$. Using a similar approach as above, we obtain 
\begin{align}\label{eqn:final-4}
p_n(t) &=  \sum_{j=0}^{\infty} \pr\{Q(S_k + T_k^{\text{A}})=j\} \cdot \pr\{Q(t) = n\,|\, Q(S_k+T_k^{\text{A}}) = j\} \nonumber \\
&= \sum_{j=0}^{\infty } p_j(S_k+T_k^{\text{A}}) \cdot G_{j,n}(\Delta t; \lambda_k^{\text{I}}, \mu_k),
\end{align}
where $\Delta t = t - S_k-T_k^{\text{A}}$ is the elapsed time from the beginning of the $k$th inactive subinterval. Combining (\ref{eqn:final-3}), (\ref{eqn:final-4}), and Lemma~3, we have the following theorem for the transient solution $p_n(t)$ of the problem (\ref{eqn:governing}). 

\smallskip
\noindent \textbf{Theorem~1 (Transient solution at an arbitrary point in time).} \\
Let $p(t) = [p_n(t), n\geq 0]$. Then, for $t\geq 0$, we have
\begin{align*}
p(t) =  \begin{cases}
p(S_k)\cdot G_k^{\text{A}}(t-S_k ) & \text{if}\,\, t \in [S_k,  S_k + T_k^{\text{A}}),\\
p(S_k+T_k^{\text{A}})\cdot G_k^{\text{I}}(t-S_k - T_k^{\text{A}}) &\text{if}\,\, t \in [S_k + T_k^{\text{A}}, S_{k+1}),
\end{cases}
\end{align*}
where $p(S_k)$ and $p(S_k+T_k^{\text{A}})$ are given in Lemma~3, and $G_s^{\text{A}}(\cdot)$ and $G_s^{\text{I}}(\cdot)$ are matrices defined by 
\begin{align*}
    G_s^{\text{A}}(x) & \deq \big[G_{i,j}(x; \lambda_s^{\text{A}}, \mu_s)\big]_{i,j\geq 0},\\
    G_s^{\text{I}}(x) & \deq \big[G_{i,j}(x; \lambda_s^{\text{I}}, \mu_s)\big]_{i,j\geq 0}. 
\end{align*}
\noindent \textbf{Proof.}  The formulas (\ref{eqn:final-3}) and (\ref{eqn:final-4}) hold for all $n=0,1,2,\ldots$. Hence, representing them in matrix form gives Theorem~1. \hfill $\blacksquare$
\smallskip

\noindent \textbf{Remark 2.} Our framework works for any a priori given sequence of active and inactive durations $\{(T_k^{\text{A}},T_k^{\text{I}}), k \geq 1 \}$ even if the values of $T_k^{\text{A}}$ and $T_k^{\text{I}}$ are all different across $k$. In the case when $T_k^{\text{A}}$ and $T_k^{\text{I}}$ are independent exponential random variables, Berg and Groenendijk~\cite{transient1991} presented a numerical method for computing the probability generating function of the queue length as a function
of time in an M/M/1 queue with regularly changing arrival and
service intensities. Hence, the method in~\cite{transient1991} requires an additional step of inverting probability generating functions  in order to find the queue length distribution.

\smallskip
\noindent \textbf{Remark 3.} We demonstrate the extensibility of our analytical framework to the case when  $T_k^{\text{A}}$ and $T_k^{\text{I}}$ are random variables that are independent but non-identically distributed across~$k$. As for the transient solution at each switching point, a recursive relation similar to (\ref{eqn:initial-3}) and (\ref{eqn:initial-4}) holds as follows:
\begin{align}\label{eqn:random1}
\begin{split}
    I_{k,n} &=  \sum_{j=0}^{\infty}A_{k,j} \cdot \E[G_{j,n}(T_k^{\text{A}}; \lambda_k^{\text{A}}, \mu_k)], \\
A_{k+1,n} 
&= \sum_{j=0}^{\infty} I_{k,j}\cdot \E[G_{j,n}(T_{k}^{\text{I}}; \lambda_{k}^{\text{I}}, \mu_{k})],
\end{split}
\end{align}
where the expectation is taken to the function $G_{j,n}(T; \lambda, \mu)$ with respect to the random variable $T$. The proof of (\ref{eqn:random1}) is given in Appendix~A.

Concerning the violation probability, we define
\begin{align*}
   V_k^{\text{A}} &\deq \frac{\E[\int_{S_k}^{S_{k}+T_k^{\text{A}}} 1_{\{Q(s) > q_{\text{th}}\}} \, \mathrm{d}s ]}{\E[T_k^{\text{A}}]}, \\
 V_k^{\text{I}} &\deq \frac{\E[\int_{S_{k}+T_k^{\text{A}}}^{S_{k+1}} 1_{\{Q(s) > q_{\text{th}}\}} \, \mathrm{d}s ]}{\E[T_k^{\text{I}}]}, 
\end{align*}
which represent the expected violation probabilities in the $k$th active and inactive subintervals, respectively. Note that in the derivation of $p_n(t)$ for an arbitrary time $t\geq 0$, we decomposed $t$ as $t = S_k + \Delta t$ or $t  = S_k + T_k^{\text{A}} + \Delta t$ (i.e., the sum of the switching point and the elapsed time from the switching point), and express $p_n(t)$ in terms of $A_{k,j}\cdot G_{j,n}(\Delta t; \lambda_k^{\text{A}}, \mu_k)$ or $I_{k,j}\cdot G_{j,n}(\Delta t; \lambda_k^{\text{I}}, \mu_k)$ (see (\ref{eqn:final-3}) and (\ref{eqn:final-4})). Applying this approach again, we obtain 
\begin{align}\label{eqn:random2}
\begin{split}
       V_k^{\text{A}} 
   &= 1-\frac{1}{\E[T_k^{\text{A}}]}\sum_{n = 0}^{q_{\text{th}}} \sum_{j=0}^{\infty} A_{k,j}\cdot \E[\int_{0}^{T_k^{\text{A}}} G_{j,n}(t;\lambda_k^{\text{A}},\mu_k)\, \mathrm{d}t ],\\
      V_k^{\text{I}} 
  &= 1-\frac{1}{\E[T_k^{\text{I}}]}\sum_{n = 0}^{q_{\text{th}}} \sum_{j=0}^{\infty} I_{k,j}\cdot \E[\int_{0}^{T_k^{\text{I}}} G_{j,n}(t;\lambda_k^{\text{I}},\mu_k)\, \mathrm{d} t ].
  \end{split}
\end{align}
The proof of (\ref{eqn:random2}) is given in Appendix~B. Then, the time-averaged violation probability from the beginning (i.e., $t=0$) to the end of the $k$th interval (i.e., $t=S_{k+1}$) can be computed using the metrics $V_k^{\text{A}}$ and $V_k^{\text{I}}$ as follows:
\begin{align*}
\bar{V}_k &\deq  \frac{\E[\int_{0}^{S_{k+1}} 1_{\{Q(s) > q_{\text{th}}\}} \, \mathrm{d}s ]}{\E[S_{k+1}]}  \\
&= \frac{1}{\E[S_{k+1}]}\sum_{n=1}^{k}\E[\int_{S_n}^{S_{n+1}} 1_{\{Q(s) > q_{\text{th}}\}} \, \mathrm{d}s ] \\
&= \frac{1}{\sum_{n=1}^{k}(\E[T_n^{\text{A}}] + \E[T_n^{\text{I}}])}\sum_{n=1}^{k}(V_n^{\text{A}}\cdot \E[T_n^{\text{A}}] + V_n^{\text{I}}\cdot \E[T_n^{\text{I}}]).
\end{align*}
\pdfoutput=1
\begin{figure*}[!t]
	\centering
	\subfigure[Settings for an on-off traffic model ($\lambda^\text{on}_k=10$, $\lambda^\text{off}_k=0$, $\mu_k = 5$, $T^\text{on}_k = T^\text{off}_k = 0.5$)]{\includegraphics[width=0.32\textwidth]{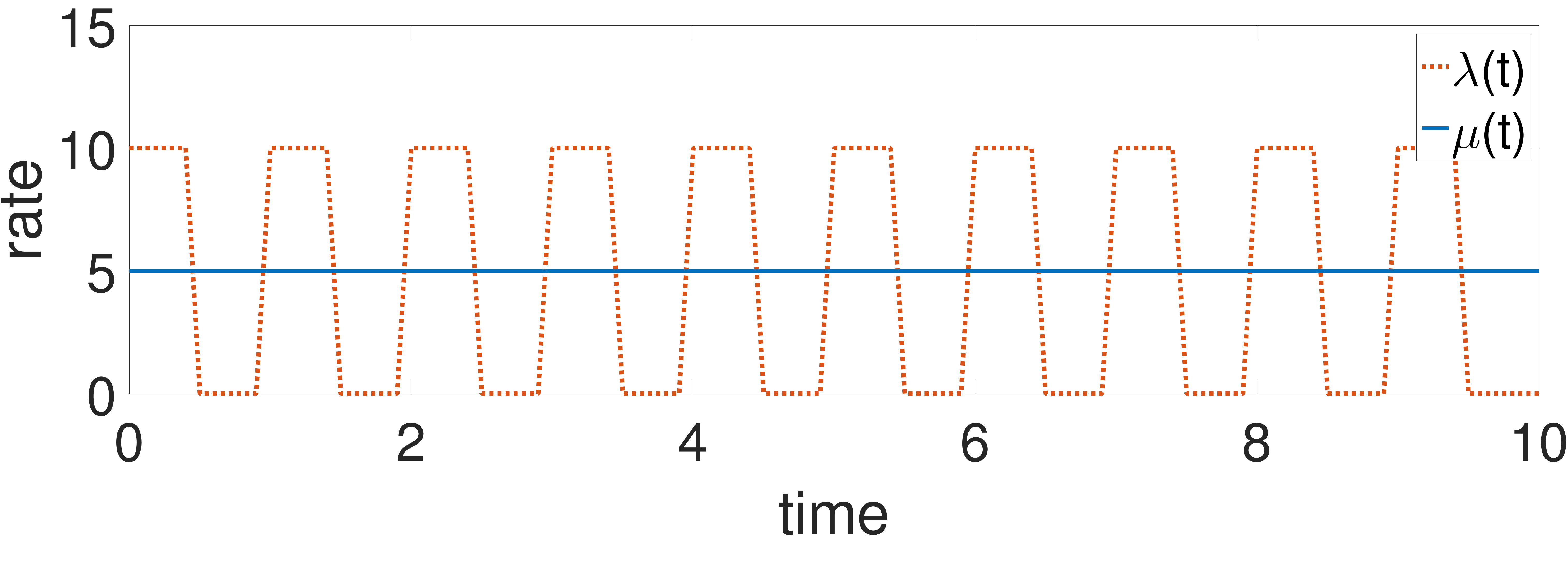}}
	\subfigure[Settings for a high-low traffic  model ($\lambda^\text{H}_{k}=k+4$, $\lambda^\text{L}_{k}= 2 - (k \mod 2)$, $\mu_k = 5$, $T^\text{H}_k = T^\text{L}_k = 0.5$)]{\includegraphics[width=0.32\textwidth]{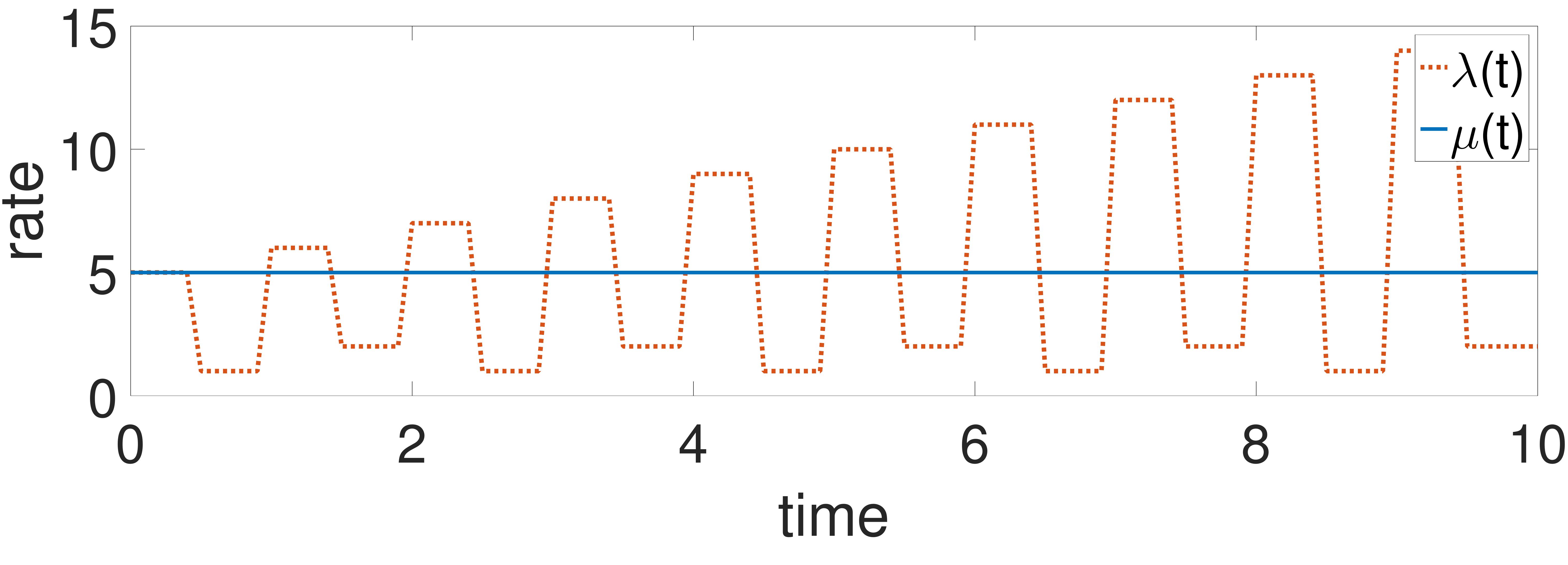}}
	\subfigure[Settings for a restless traffic model ($\lambda^\text{A}_k =10-|k-6|$, $\mu_k = |k-6|+5$, $T^\text{A}_k = 1$, $T^\text{I}_k = 0$)]{\includegraphics[width=0.32\textwidth]{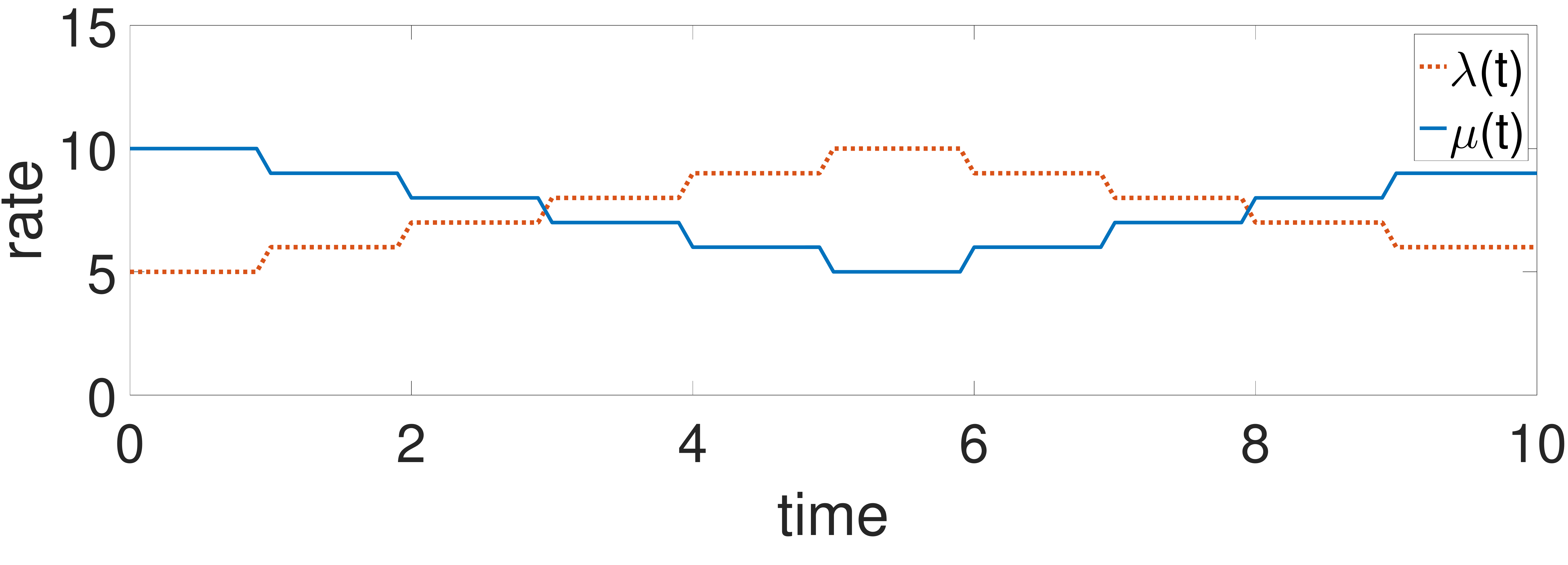}}\\
	
	\subfigure[$V(t)$ from the on-off traffic model ($q_\text{th}=5$)]{\includegraphics[width=0.32\textwidth]{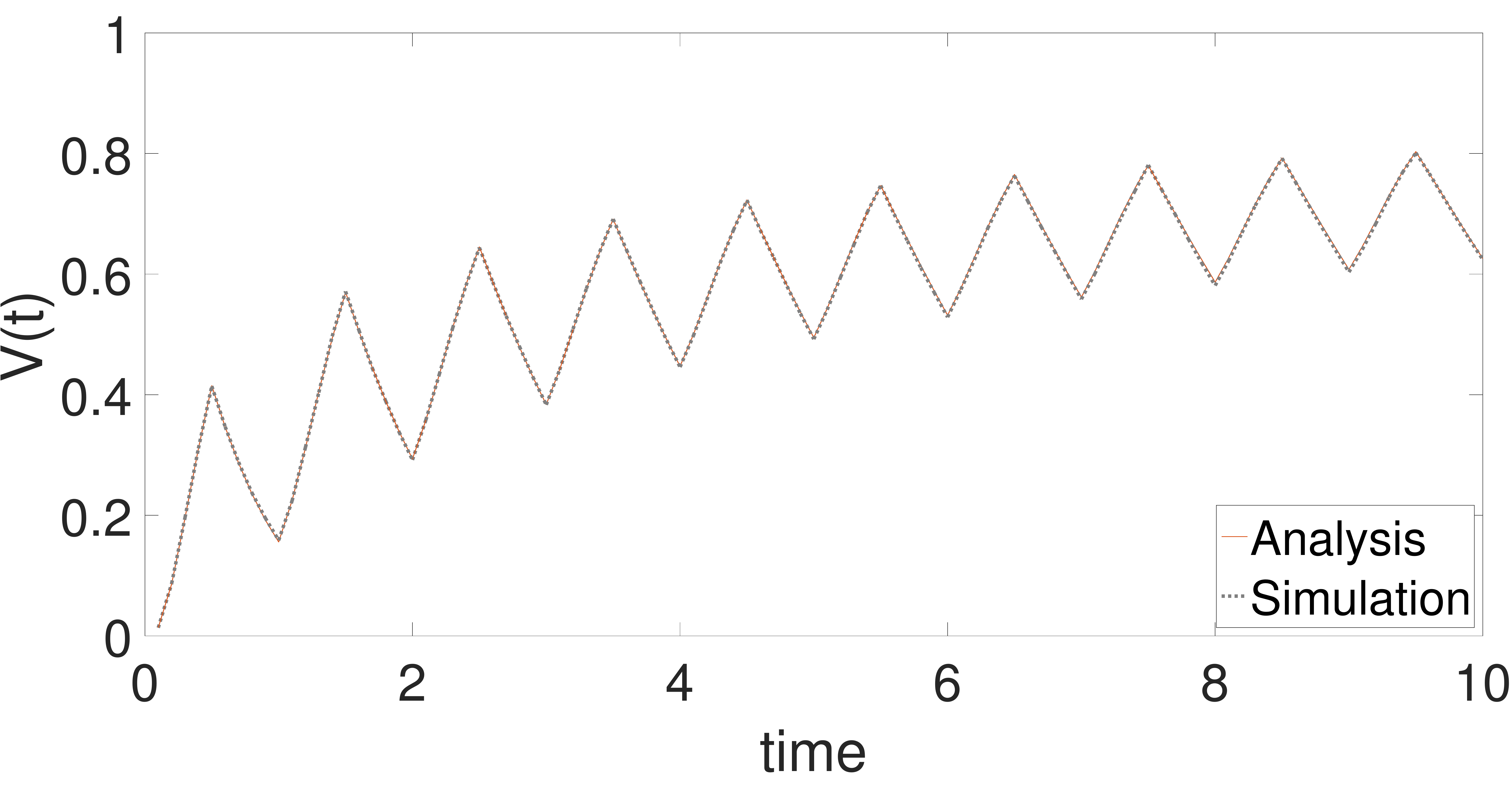}}
	\subfigure[$V(t)$ from the high-low traffic model ($q_\text{th}=5$)]{\includegraphics[width=0.32\textwidth]{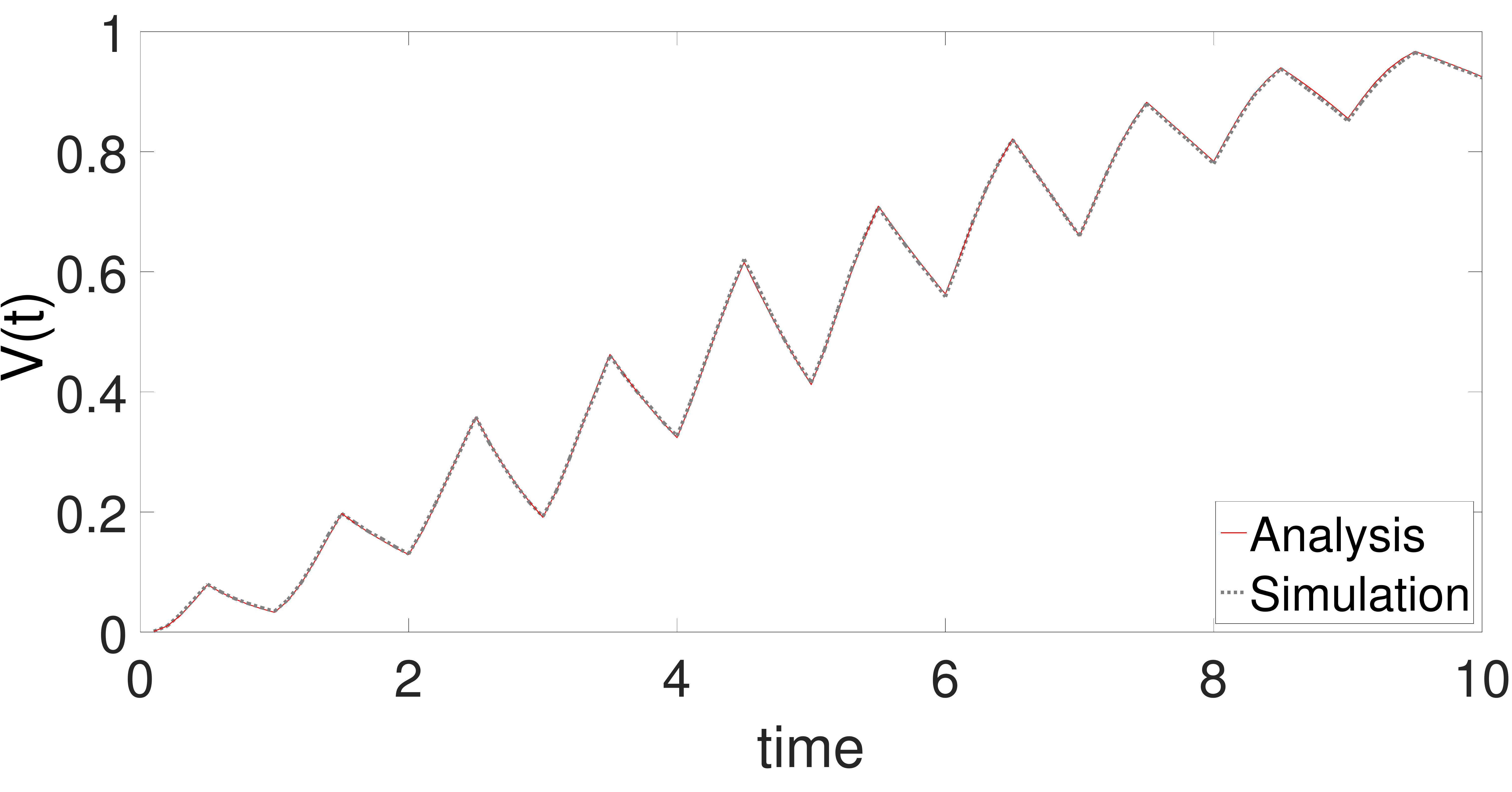}}
	\subfigure[$V(t)$ from the restless traffic model ($q_\text{th}=5$)]{\includegraphics[width=0.32\textwidth]{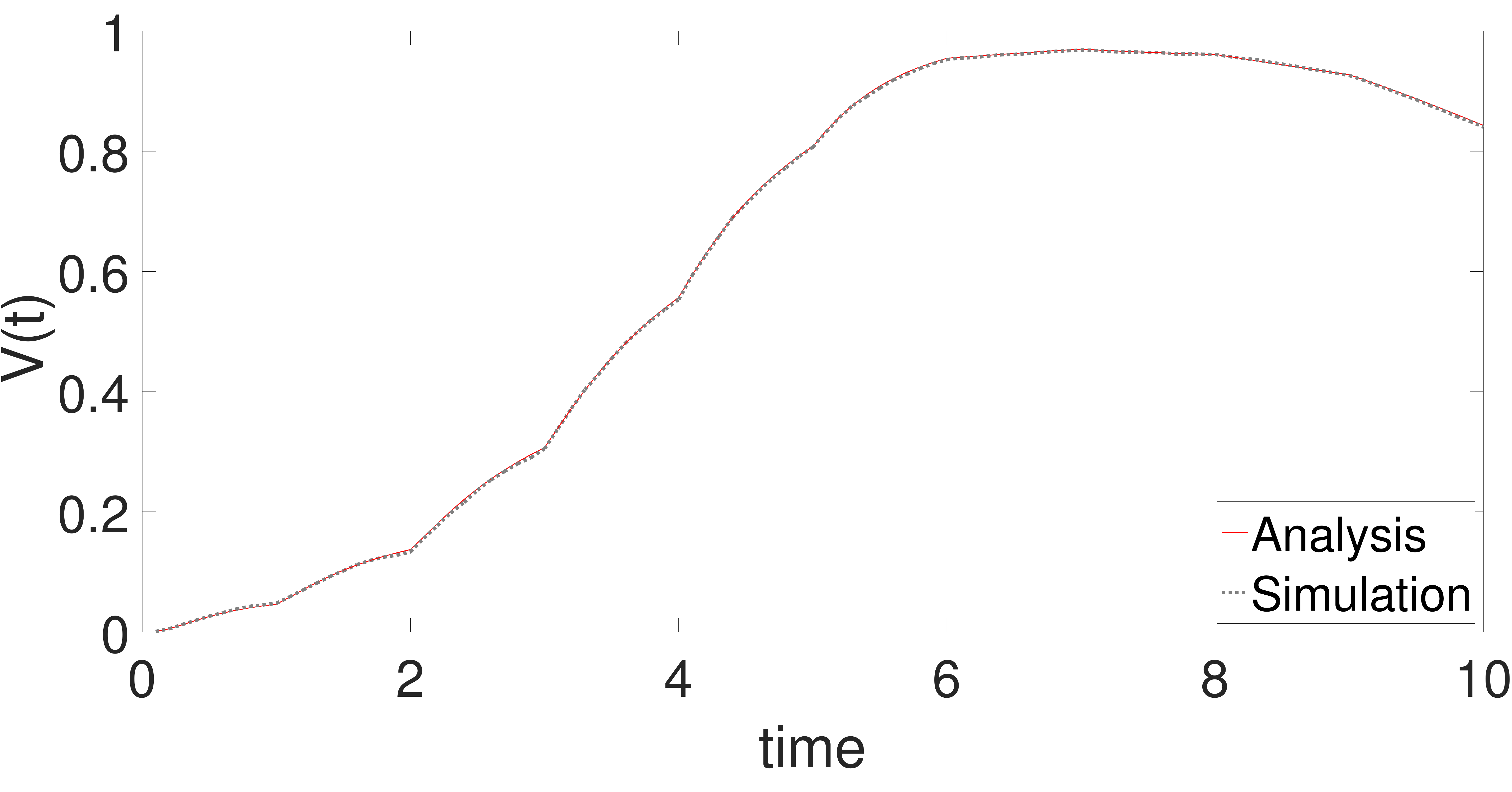}}
    \caption{The violation probability over 10 intervals with (d) on-off traffic model, (e) high-low traffic model, and (f) restless traffic model, where the duration of each interval is 1 and $q_\text{th}$ is 5. Their time-varying arrival and service rates are depicted in (a), (b), and (c), respectively.}	
    \label{fig:vt-models}
\end{figure*}

\section{Performance Evaluation}
\label{sec:evaluation}

In this section, we first validate our analytical framework in comparison with simulation results through several example scenarios and demonstrate its capability in predicting the queue behavior under various arrival and service rate patterns. 

\smallskip
\subsection{Validation of Our Framework}
For the validation of our framework, we focus on analyzing one link of a network, whose traffic input and output patterns follow one of our traffic models described in Section~\ref{sec:system}. We compare $V(t)$, the violation probability at time $t$, for a given queue length threshold between the results from our analysis and \textcolor{black}{those} from simulations performed in MATLAB. In order to make the simulation results to be statistically valid, for each test we take the average  over  a hundred thousand simulation runs (i.e., $10^5$ runs). For all validations, we set the number of intervals to evaluate as 10 and use common settings for the following parameters, \textcolor{black}{unless otherwise mentioned}: \textcolor{black}{$q_\text{th} = 5, T_k^\text{A} = T_k^\text{I} = 0.5\,(k=1,2,\ldots,10)$}. 

\smallskip
\noindent{\bf{Under an On-Off Traffic Model:}}
To emphasize the transient queue behavior in the on-off traffic model, we evaluate $V(t)$ while we opt to maintain $\lambda_k^\text{on}, \lambda_k^\text{off}$, and $\mu_k$ as 10, 0, and 5, respectively, as shown in Figure~\ref{fig:vt-models}~(a). Figure~\ref{fig:vt-models}~(d) compares the result from our analytical framework with that from simulations and confirms that they closely match each other. As it is expected, both results show that the violation probability increases as the intervals proceed due to heavy packet arrivals, but further show that the queue does not explode during 10 intervals and takes up and down as on and off subintervals switch. This implies that even if a short-lived low-latency service generates packets intensively in this manner, there is still a chance to have the overall latency violation managed within a certain level. This is unforeseeable from any steady-state analysis. 

\smallskip
\noindent{\bf{Under a High-Low Traffic Model:}}
With a high-low traffic model, we test a more complicated scenario where background traffic persists even through inactive subintervals and $\lambda_k^\text{H}$ keeps increasing through active subintervals. To be more specific, we set $\lambda^\text{H}_{k}=k+4$ and $\lambda^\text{L}_{k}= 2 - (k \mod 2)$ for the $k$th interval, while $\mu_k = 5$ is kept through all intervals as depicted in Figure~\ref{fig:vt-models}~(b). Figure~\ref{fig:vt-models}~(e) shows that our framework well predicts the progressive expansion of queue length. Our framework also well captures the contraction behavior of the queue during the inactive subintervals where the arrival rate stays below the service rate.


\smallskip
\noindent{\bf{Under a Restless Traffic Model:}}
In the restless traffic model, we consider an entire interval to be active, and thus inactive subintervals do not exist. Therefore, $T_k^\text{A} = 1$ and $T_k^\text{I} = 0$ hold. To have a differentiated scenario, we assume stronger dynamics in both arrival and service rates. The arrival rate is set to increase and then to decrease as $\lambda^\text{A}_{k}=10-|k-6|$ and the service rate is also set to change as $\mu_{k}=|k-6|+5$ as in Figure~\ref{fig:vt-models}~(c). With such dynamics, Figure~\ref{fig:vt-models}~(f) shows that the queue quickly fills up until $k=8$ where the arrival rate becomes higher than the service rate. Figure~\ref{fig:vt-models}~(f) also shows that from the interval $k=8$ when the service rate surpasses the arrival rate, the violation probability diminishes.  


\subsection{Impact of $q_\text{th}$}
\begin{figure*}[t!]
    \centering
	\subfigure[$\bar{V}$ for $q_{\text{th}} = 5$]{\centering \includegraphics[width=0.32\textwidth]{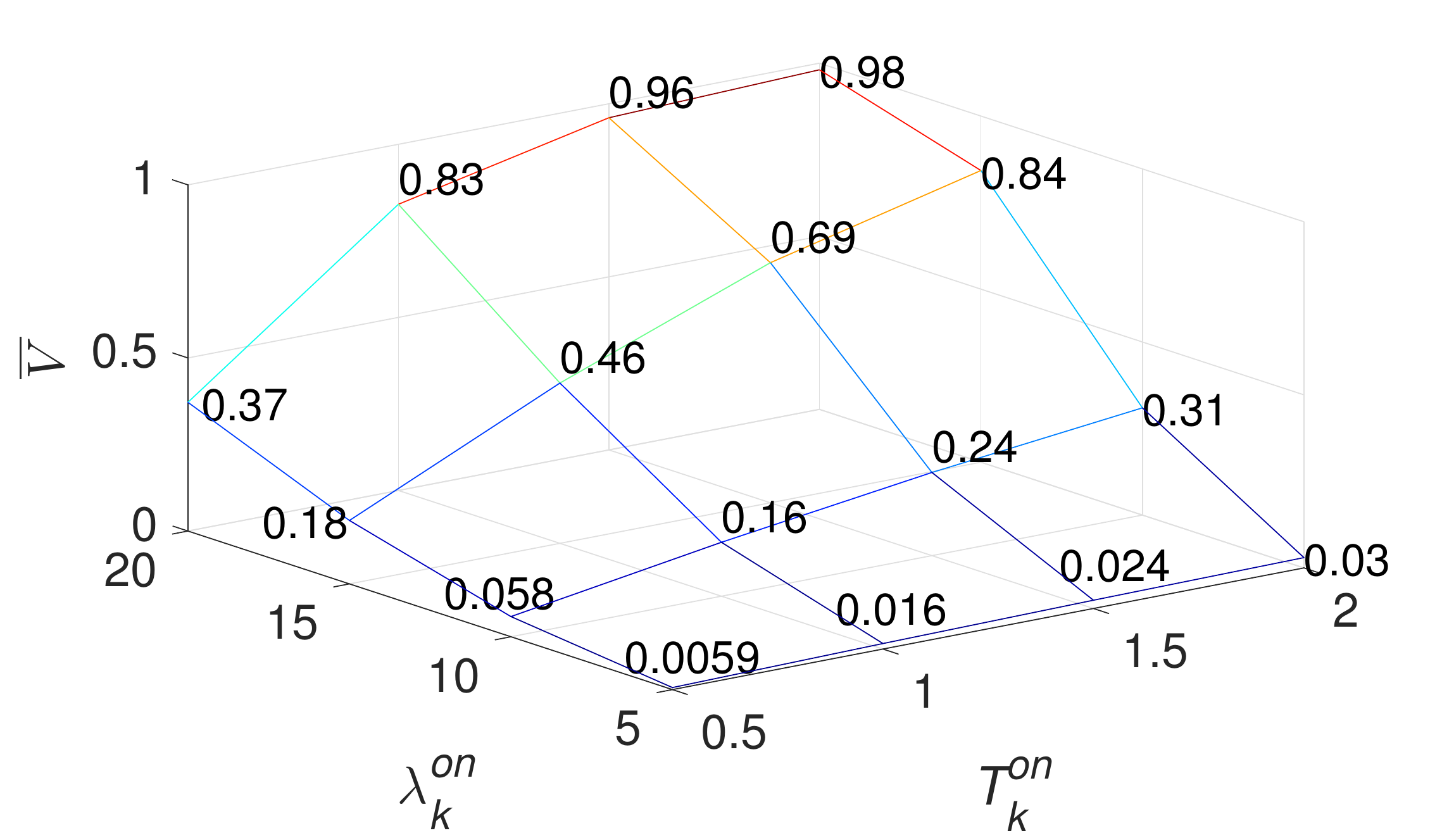}}
    \subfigure[$\bar{V}$ for $q_{\text{th}} = 10$]{\centering \includegraphics[width=0.32\textwidth]{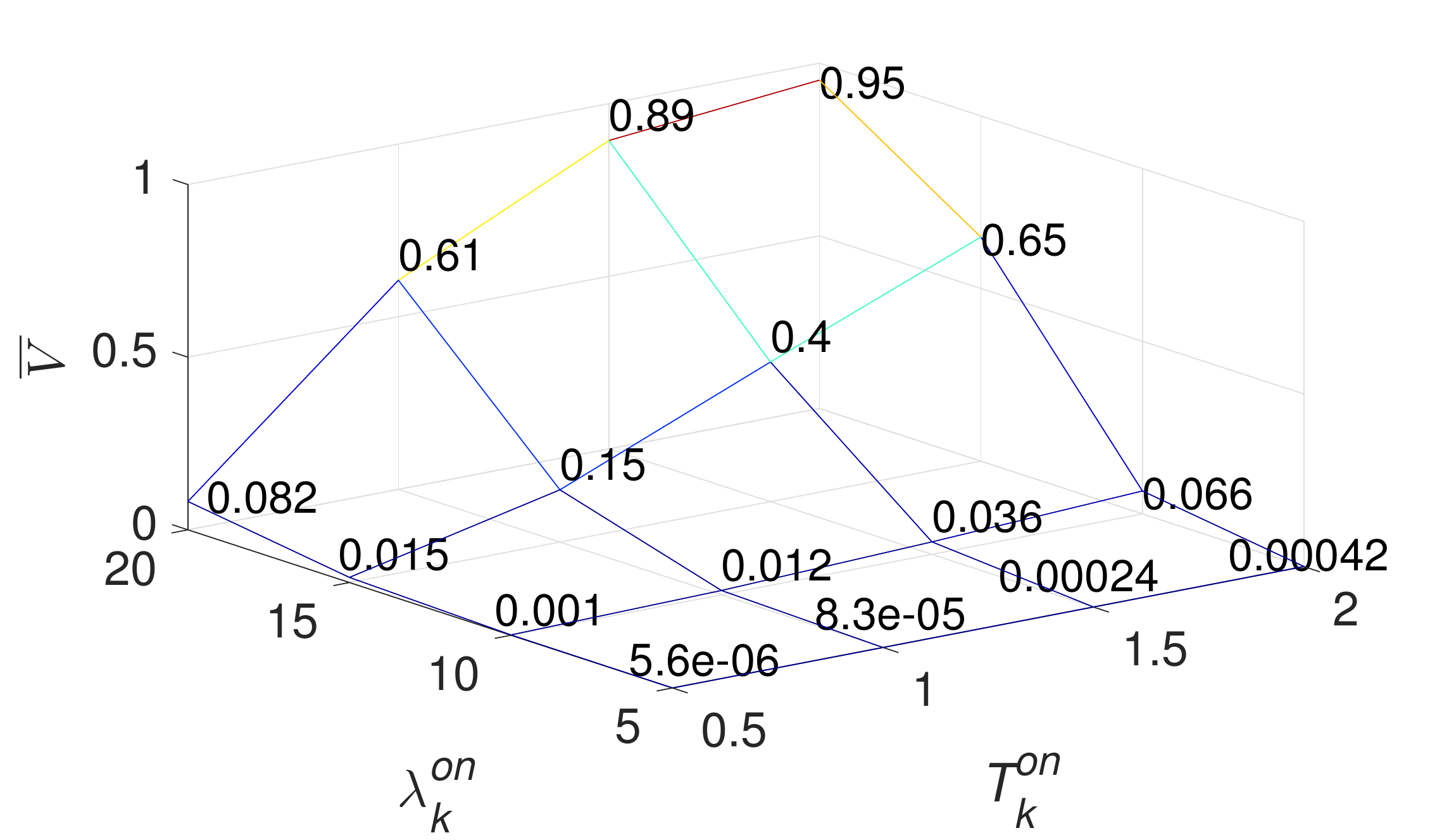}}
    \subfigure[$\bar{V}$ for $q_{\text{th}} = 15$]{\centering \includegraphics[width=0.32\textwidth]{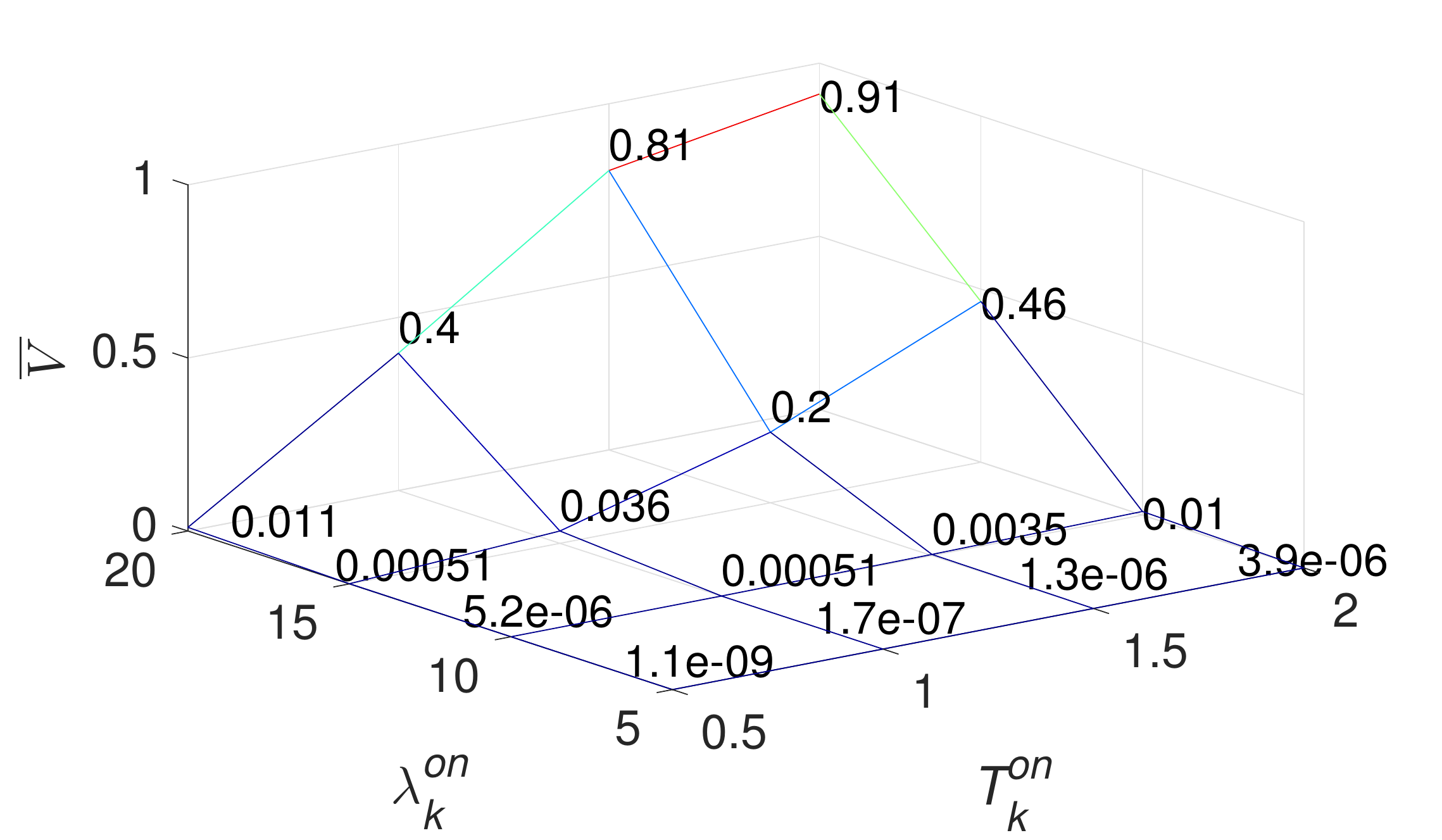}}
    \vspace{-0.3cm}
	\caption{The \textcolor{black} {time-averaged} violation probability ($\bar{V}$) for three different $q_\text{th}$ values under various \textcolor{black}{$\lambda^\text{on}_k$ and $T^\text{on}_k$} settings.}
	\label{fig:3d-qth}
	\vspace{-0.5cm}
\end{figure*}

We now move our focus from $V(t)$ to $\bar{V}$, the time-averaged violation probability through the entire intervals and test the impact of $q_\text{th}$ on $\bar{V}$. 
For simplicity, we opt to use on-off traffic model in which the arrival rate remains zero while being at inactive subintervals. For this, we fix $T_k^\text{off} = 1$ and vary $T_k^\text{on}$ to have either of 0.5, 1, 1.5, and 2. We also vary $\lambda_k^\text{on}$ to have either of 5, 10, 15, and 20 while the service rate $\mu_k$ always is set to stay at 5. The number of total intervals is maintained to be always 10, but notice that the total duration varies as $T_k^\text{on}$ varies. Figures~\ref{fig:3d-qth} (a), (b), and (c) show the 3D plots of $\bar{V}$ for $q_{\text{th}}$ being 5, 10, and 15, respectively. It is intuitive to expect that a higher $q_{\text{th}}$ leads to a smaller $\bar{V}$, but how much is challenging to be answered. Our framework is fully capable of answering how much. Given that assessing the time-averaged violation probability for a specific $q_{\text{th}}$ is directly connected to quantifying the portion of packets of a session that fails to meet the corresponding latency bound, our framework is the first of its kind that can guide the required FEC (forward error correction) coding level for a latency-critical session to be successfully decodable with no retransmissions. To this end, our framework can be useful to the low-latency application or service designers.

\begin{figure*}[t!]
    \centering
	\subfigure[$\bar{V}$ over 20 intervals]{\centering \includegraphics[width=0.32\textwidth]{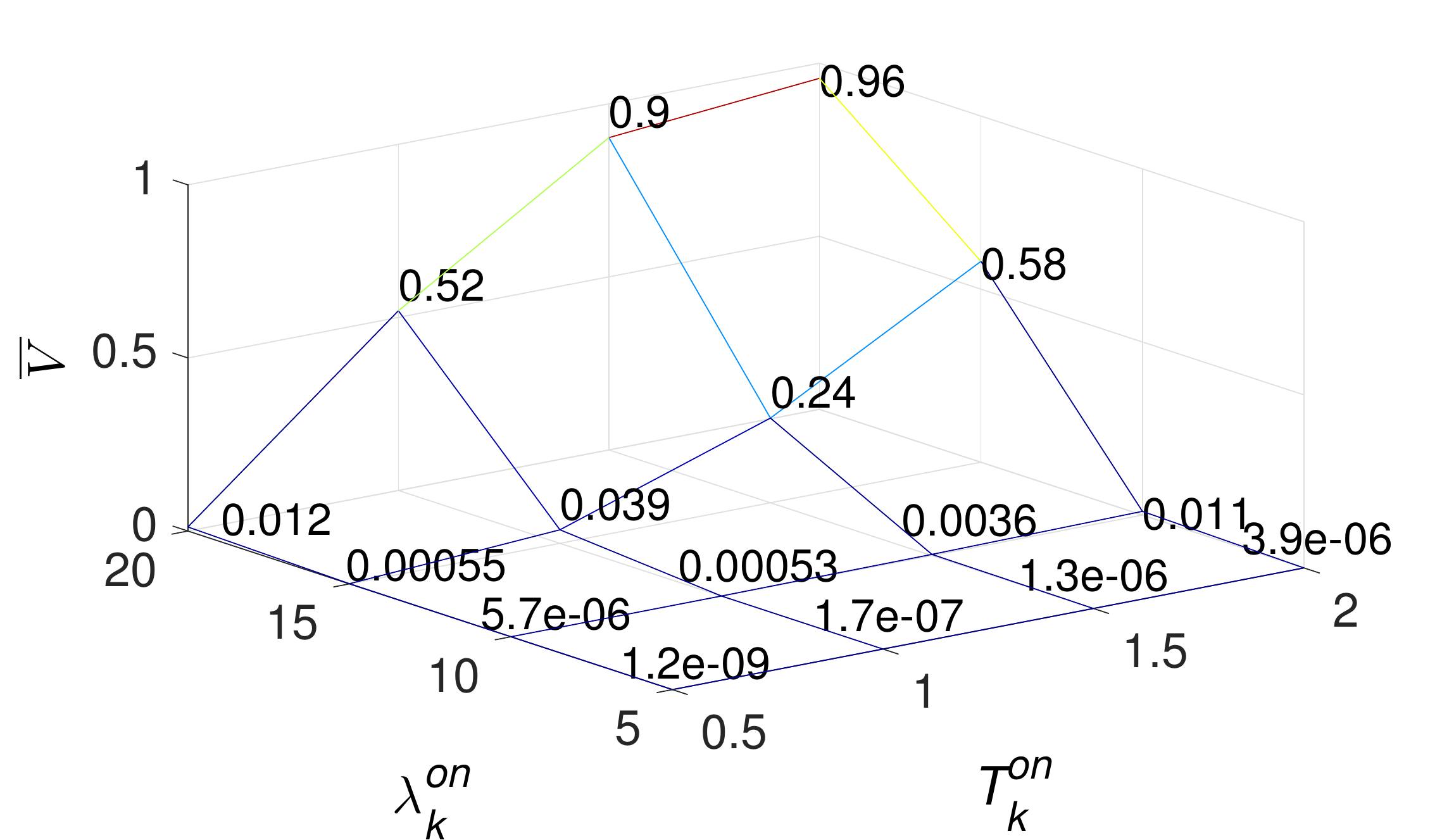}}
    \subfigure[$\bar{V}$ over 30 intervals]{\centering \includegraphics[width=0.32\textwidth]{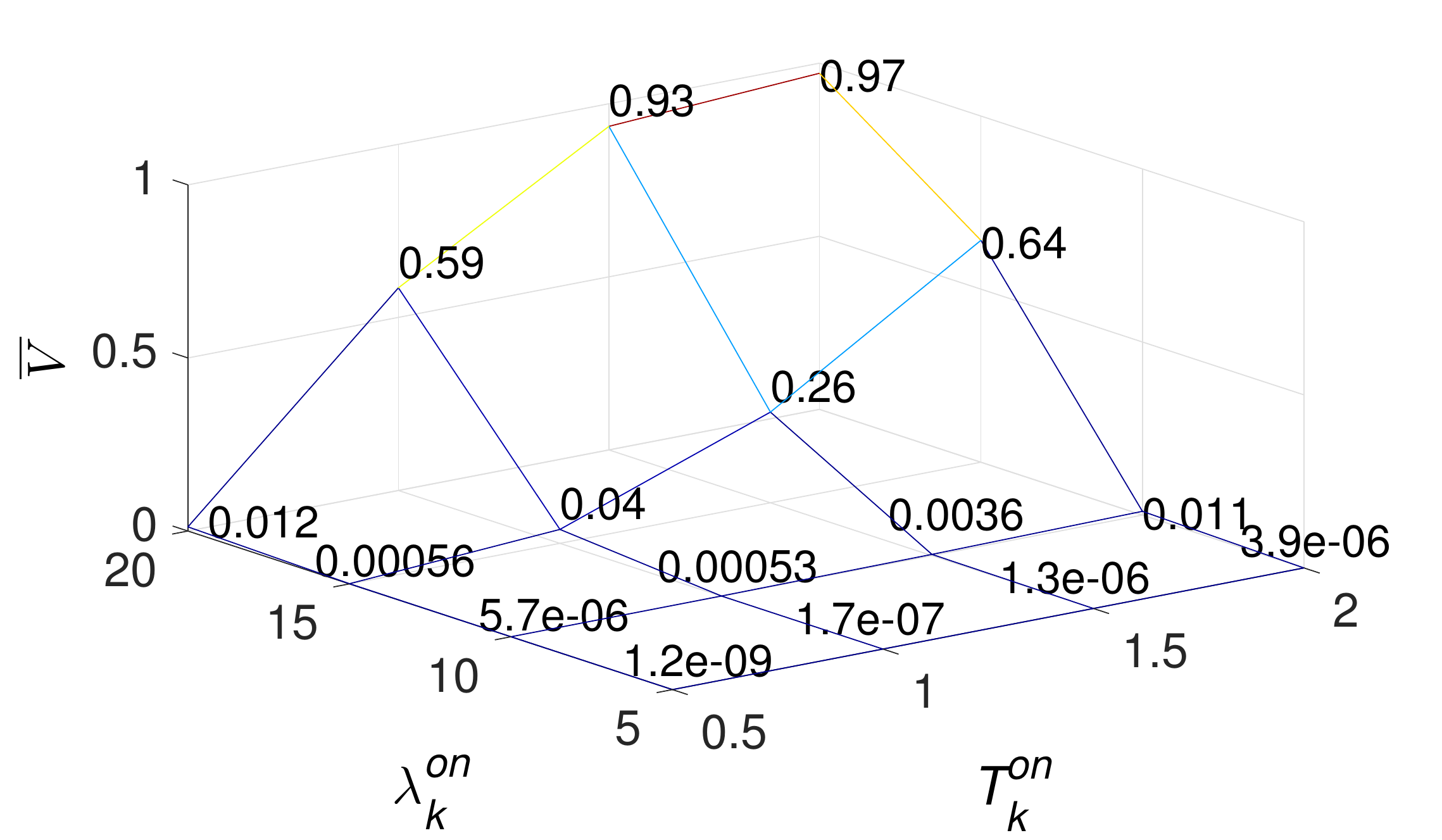}}
    \subfigure[$\bar{V}$ over 40 intervals]{\centering \includegraphics[width=0.32\textwidth]{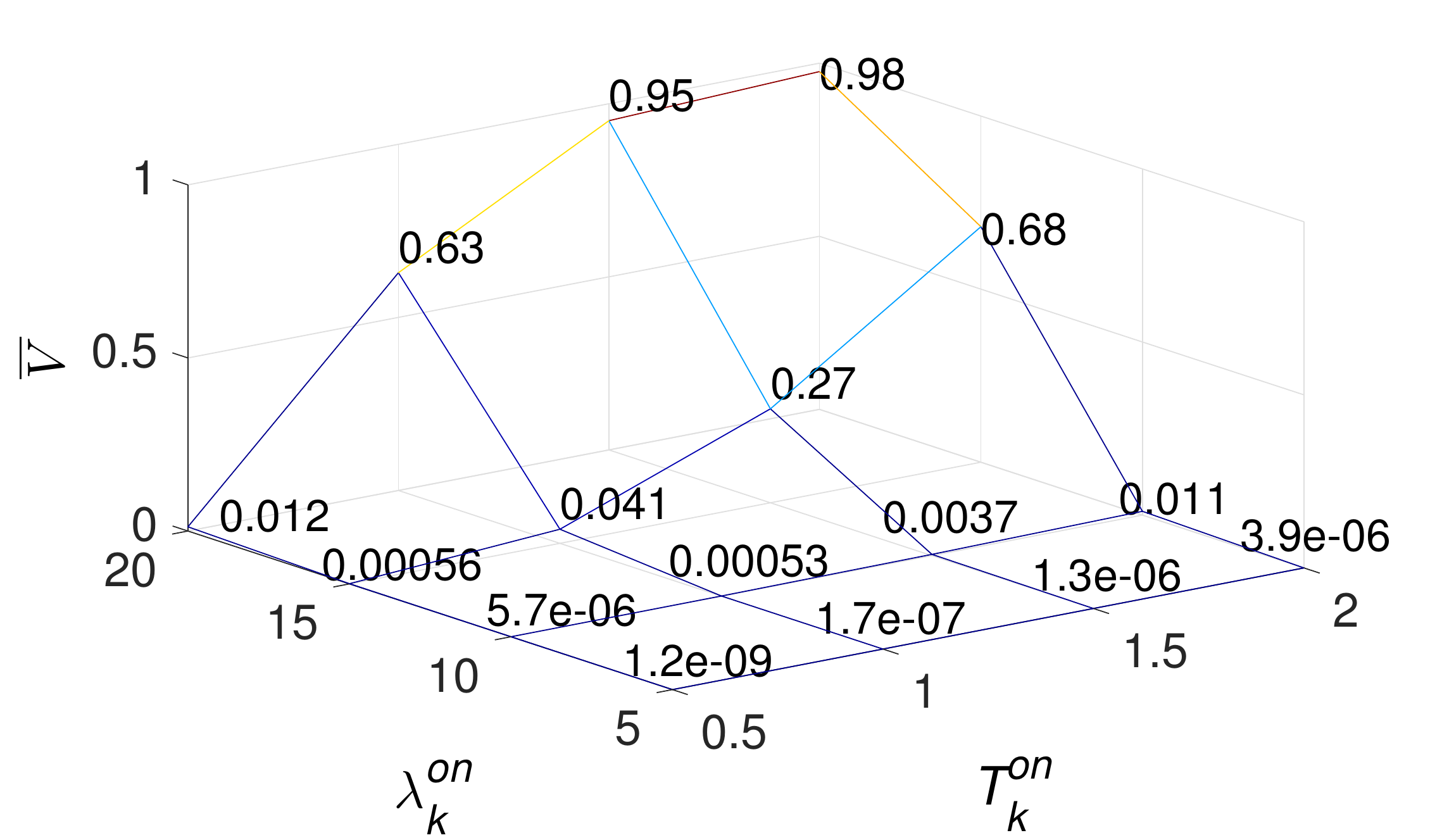}}
    \vspace{-0.3cm}
	\caption{The \textcolor{black}{time-averaged} violation probability ($\bar{V}$) for \textcolor{black}{different number of intervals} under various $\lambda^\text{on}_k$ and $T^\text{on}_k$ settings.}
	\label{fig:3d-interval}
	\vspace{-0.5cm}
\end{figure*}

\begin{figure*}[t!]
    \centering    
	\subfigure[$\bar{V}$ for decreasing $\mu_k$]{\centering \includegraphics[width=0.32\textwidth]{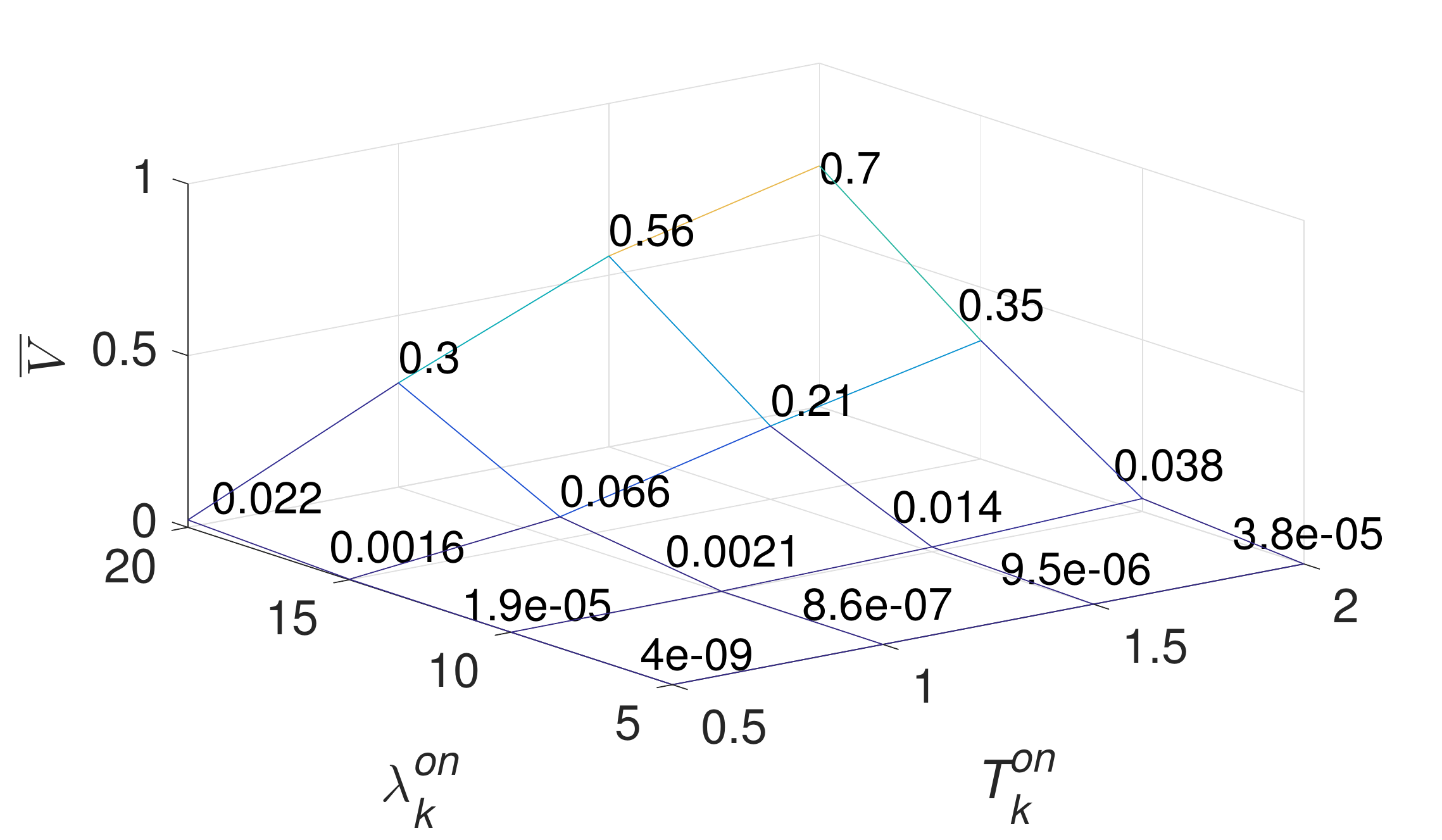}}
    \subfigure[$\bar{V}$ for increasing $\mu_k$]{\centering \includegraphics[width=0.32\textwidth]{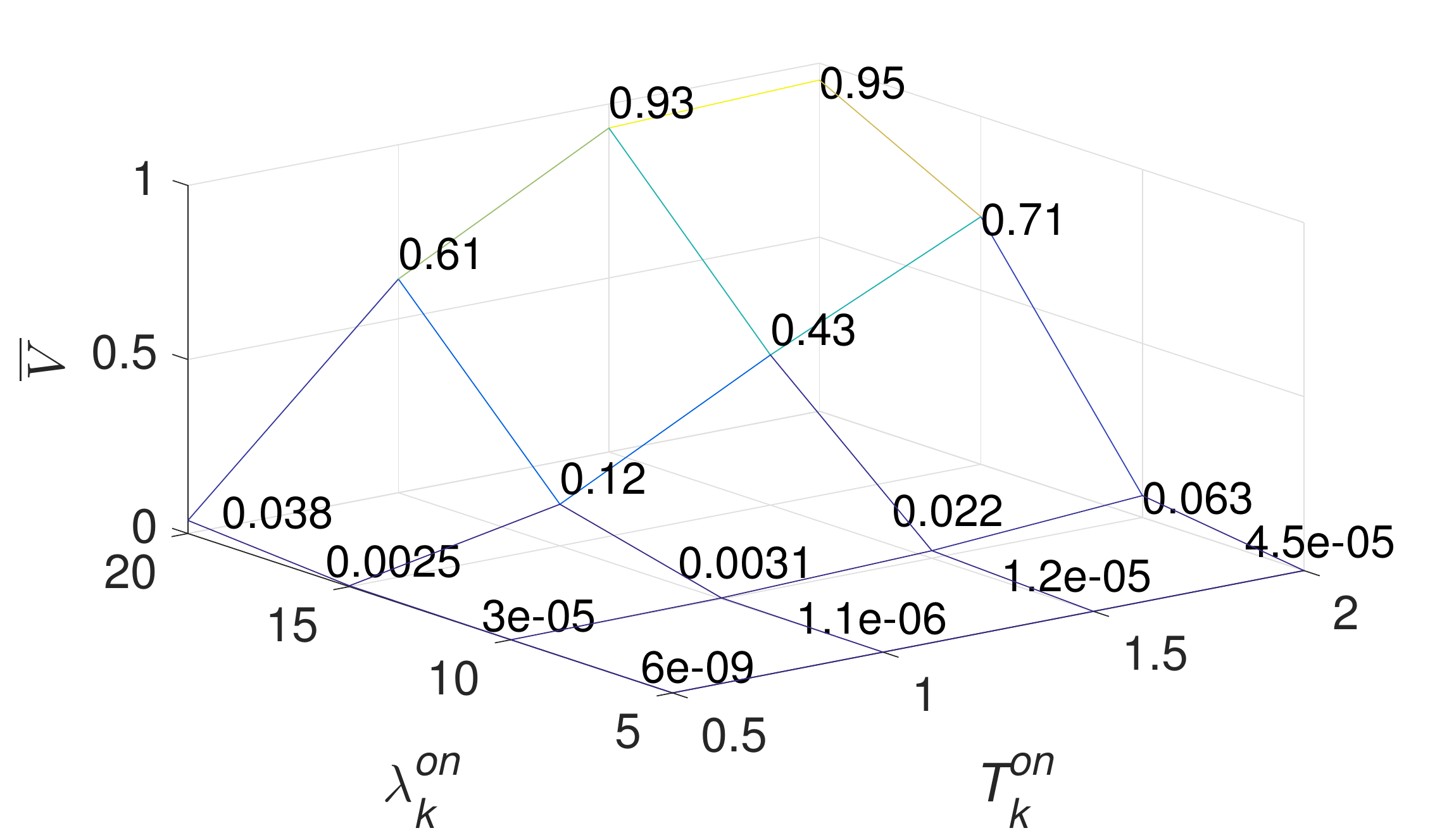}}
    \subfigure[$\bar{V}$ for decreasing and then increasing $\mu_k$]{\centering \includegraphics[width=0.32\textwidth]{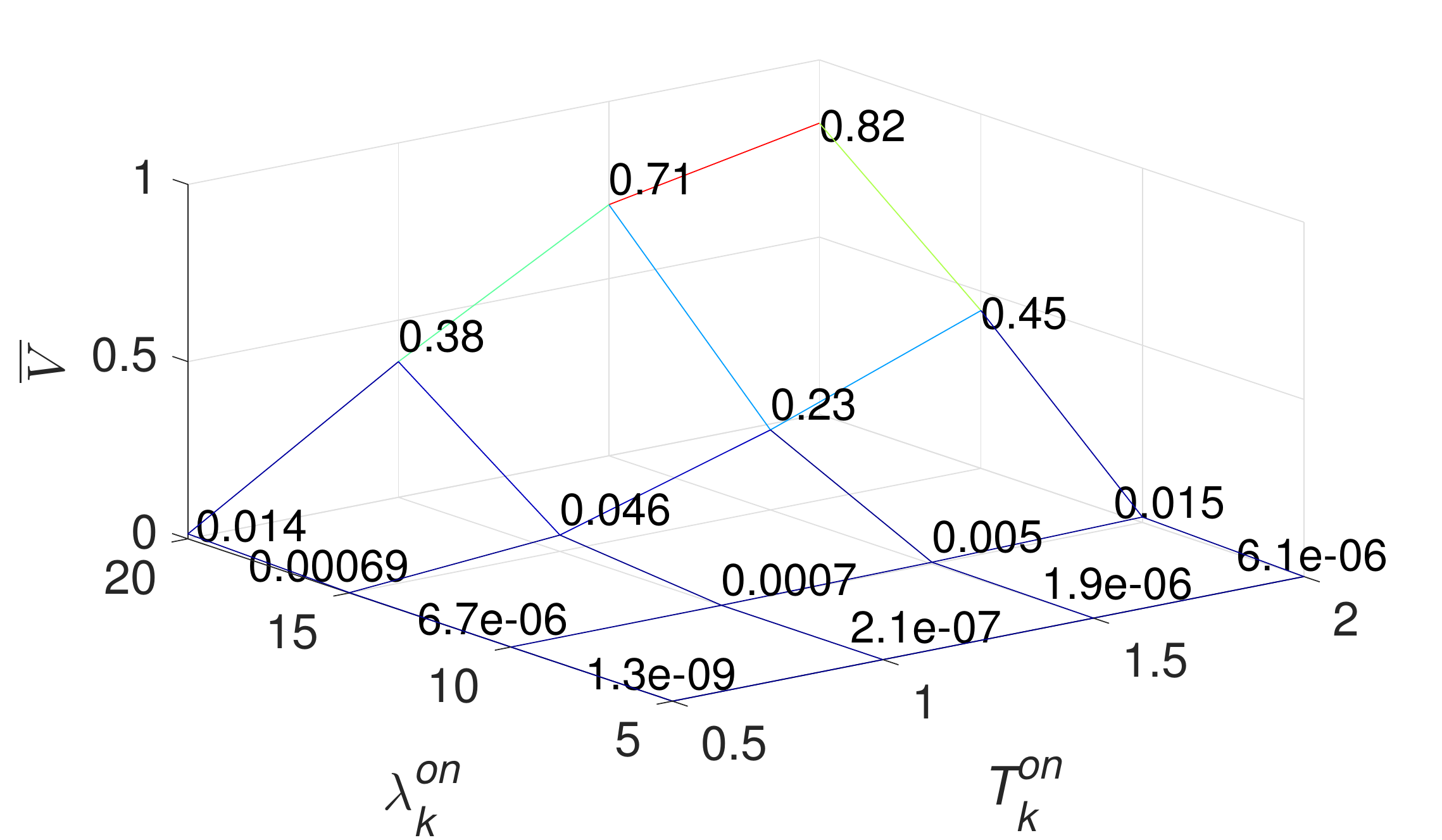}}
    \vspace{-0.3cm}
	\caption{The \textcolor{black}{time-averaged} violation probability ($\bar{V}$) for three time-varying patterns of $\mu_k$ under various $\lambda^\text{on}_k$ and $T^\text{on}_k$ settings.}
	\label{fig:3d-mu}
\end{figure*}

\subsection{Impact of the Number of Intervals}
We next study the impact of the number of intervals. It is intuitive that having the same input and output traffic patterns repeated for a longer duration of time results in saturation. However, how quickly the saturation will happen has been under-explored in conventional queueing analysis. Through Figures~\ref{fig:3d-interval} (a), (b), and (c), we show how $\bar{V}$ changes as the number of intervals lengthens as 20, 30, and 40. Figures~\ref{fig:3d-interval} (a), (b), and (c) are connected to Figure~\ref{fig:3d-qth} (c) whose number of intervals is 10 and all other parameter settings are identical. It is interesting to observe that $\bar{V}$ from $T^\text{on}_k =1.5$ and $\lambda^\text{on}_k = 15$ suddenly jumps from 0.2 to 0.24 when switching from 10 intervals to 20 intervals, while the values stay almost saturated as 0.26 and 0.27 at 30 and 40 intervals. These numbers are valuable to the designers of short-lived low-latency service sessions, such as real-time IoT communications.

\subsection{Impact of Time-Varying $\mu_k$}
We take a further step toward understanding the impact of time-varying service rate. Out of many possible scenarios, we choose to study three cases in which the service rate keeps decreasing, keeps increasing, and decreases and then increases. We think that these cases are particularly intriguing as these patterns of $\mu_k$ may be observed when the mobility is involved (e.g., moving toward a worse or a better channel). Under the same combinations of $T^\text{on}_k$ and $\lambda^\text{on}_k$ as in Figure~\ref{fig:3d-qth} with $q_{\text{th}} = 15$, we vary the service rate during 10 intervals as follows: (a) $\mu_k = 16-k$, (b) $\mu_k = k+5$, and (c) $\mu_k = 8 + |k - 6|$. Note that the average values of $\mu_k$ in three cases are the same to each other as 10.5. Figures~\ref{fig:3d-mu} (a), (b), and (c) show that although the average service rate is the same, the time-averaged violation probabilities can be very different. Especially, it is interesting to observe that having increasing $\mu_k$ suffers the most from the violation. This can be understood as that an early pileup from the small service rate in the beginning affects negatively to the queue length and this negative impact lasts long over multiple intervals. As demonstrated here, the use of our framework for various patterns of time-varying $\mu_k$ can guide a low-latency service to become significantly more resilient to service rate variation especially caused by  mobility.

\section{Conclusion}
\label{sec:conclusion}

Our study on the transient analysis of queueing under time-varying arrival and service rates lets us demonstrate how much useful this analysis can be. Our framework validated through representative scenarios in comparison with simulations is shown to provide unique predictions on the violation probability over time for a given queue length threshold at diverse dynamic networking environments. We expect that the designers of emerging low-latency services demanding their traffic to be strongly latency-bounded will heavily rely on our framework in enabling the services. With our framework, they can test the impact of design factors such as traffic shapes, encoding methods, and error correction schemes, and get guided to determine the most appropriate combination. 
\section*{Appendix A}
By the law of total probability, we have
\begin{align}\label{app:initial-1}
I_{k,n}&= \sum_{j=0}^{\infty}  A_{k,j} \cdot \pr\{Q(S_k+T_k^{\text{A}}) = n\,|\, Q(S_k) = j\}.
\end{align}
From Lemma~2~(i), we further have
\begin{align}\label{app:initial-2}
\pr\{Q(S_k\!+\!T_k^{\text{A}})\! =\! n|Q(S_k)\! =\!\ j\}&\! =\! \!\int_{0}^{\infty}\!\! \pr\{Q(S_k\!+\!t) \!=\! n| Q(S_k) \!=\! j\}  \mathrm{d}F_k^{\text{A}}(t) \nonumber \\
& =  \int_{0}^{\infty} G_{j,n}(t; \lambda_k^{\text{A}}, \mu_k)\,\mathrm{d}F_k^{\text{A}}(t) \nonumber \\
&= \E[ G_{j,n}(T_k^{\text{A}}; \lambda_k^{\text{A}}, \mu_k)],
\end{align}
where $F_k^{\text{A}}(t) \deq \pr\{T_k^{\text{A}}\leq t\}$.
Substituting (\ref{eqn:initial-2}) into (\ref{eqn:initial-1}) proves the first equality in (\ref{eqn:random1}). Following a similar approach, we can prove the second equality in (\ref{eqn:random1}). Due to similarity, we omit the details. 

\section*{Appendix B}
By the change of variables, we can rewrite the numerator in the definition of $V_k^{\text{A}}$ as
\begin{align}\label{app:2:1}
    \E[\int_{S_k}^{S_{k}+T_k^{\text{A}}} 1_{\{Q(s) > q_{\text{th}}\}} \, \mathrm{d}s ] 
    & = \E[\int_{0}^{T_k^{\text{A}}} 1_{\{Q(s+S_k) > q_{\text{th}}\}} \, \mathrm{d}s ].
\end{align}
By conditioning on $T_k^{\text{A}}$, we have
\begin{align}\label{app:2:2}
 &\E[\int_{0}^{T_k^{\text{A}}} 1_{\{Q(s+S_k) > q_{\text{th}}\}} \, \mathrm{d}s ] \nonumber \\
 &= \int_{0}^{\infty} \E[\int_{0}^{t} 1_{\{Q(s+S_k) > q_{\text{th}}\}} \, \mathrm{d}s \,|\, T_k^{\text{A}} = t] \,\mathrm{d}F_k^{\text{A}}(t)\nonumber \\
    &= \int_{0}^{\infty} \int_{0}^{t} \pr\{Q(s+S_k) > q_{\text{th}}\,|\, T_k^{\text{A}} = t\} \, \mathrm{d}s \,\mathrm{d}F_k^{\text{A}}(t),
\end{align}
where $F_k^{\text{A}}(t) = \pr\{T_k^{\text{A}}\leq t\}$. For $s \in[0, t]$, we have
\begin{align*}
\pr\{Q(s+S_k) =n\,|\, T_k^{\text{A}} = t\} &= \sum_{j=0}^{\infty} A_{k,j}\cdot G_{j,n}(s;\lambda_k^{\text{A}},\mu_k).
\end{align*}
It then follows for any $s \in[0, t]$ that 
\begin{align}\label{app:2:3}
\pr\{Q(s\!+\!S_k) \!>\! q_{\text{th}}| T_k^{\text{A}} = t\} &= 1-\sum_{n=0}^{q_{\text{th}}}\sum_{j=0}^{\infty} A_{k,j}\!\cdot\! G_{j,n}(s;\lambda_k^{\text{A}},\mu_k).
\end{align}
Combining (\ref{app:2:1}), (\ref{app:2:2}) and (\ref{app:2:3}) yields
\begin{align*}
& \E[\int_{S_k}^{S_{k}+T_k^{\text{A}}} 1_{\{Q(s) > q_{\text{th}}\}} \, \mathrm{d}s ]  \\
 &= \int_{0}^{\infty} t\,\mathrm{d}F_k^{\text{A}}(t) - \sum_{n=0}^{q_{\text{th}}}\sum_{j=0}^{\infty} A_{k,j}\cdot\int_{0}^{\infty}\int_{0}^{t} G_{j,n}(s;\lambda_k^{\text{A}},\mu_k)
 \, \mathrm{d}s  \mathrm{d}F_k^{\text{A}}(t)\\
 &= \E[T_k^{\text{A}}]-\sum_{n=0}^{q_{\text{th}}}\sum_{j=0}^{\infty} A_{k,j}\cdot\E[\int_{0}^{T_k^{\text{A}}} G_{j,n}(s;\lambda_k^{\text{A}},\mu_k)
 \, \mathrm{d}s ].  
\end{align*}
Therefore, we obtain 
\begin{align*}
    V_k^{\text{A}}
   &= 1-\frac{1}{\E[T_k^{\text{A}}]}\sum_{n = 0}^{q_{\text{th}}} \sum_{j=0}^{\infty} A_{k,j}\cdot \E[\int_{0}^{T_k^{\text{A}}} G_{j,n}(t;\lambda_k^{\text{A}},\mu_k)\, \mathrm{d}t ],
\end{align*}
which proves the first equality in (\ref{eqn:random2}). Following a similar approach, we can prove the second equality in (\ref{eqn:random2}). Due to similarity, we omit the details.


\end{document}